\documentclass[10pt,journal,compsoc]{IEEEtran}

\usepackage{amssymb}
\usepackage{subcaption}
\usepackage[labelfont=bf]{caption}
\usepackage{booktabs, array, multirow}
\usepackage[algoruled]{algorithm2e}
\usepackage{algorithmic}
\usepackage{listings}
\usepackage{fancyvrb}
\fvset{fontsize=\small}
\usepackage
[acronym,smallcaps,nowarn,section,nogroupskip,nonumberlist]{glossaries}
\glsdisablehyper{}
\usepackage{threeparttable}
\usepackage{graphicx}
\usepackage{url}
\usepackage[hidelinks]{hyperref}
\usepackage{xcolor}

\ifCLASSOPTIONcompsoc
  \usepackage[nocompress]{cite}
\else
  \usepackage{cite}
\fi

\begin{document}
\title{Variational Bandwidth Auto-encoder for Hybrid Recommender Systems}

\author{Yaochen~Zhu,
        Zhenzhong~Chen,~\IEEEmembership{Senior Member,~IEEE} %<-this % stops a space
\IEEEcompsocitemizethanks{

\IEEEcompsocthanksitem This work was supported in part by National Natural Science Foundation of China under Grant No. 62036005. (Corresponding author: Zhenzhong Chen, E-mail: zzchen@ieee.org)

\IEEEcompsocthanksitem Yaochen Zhu and Zhenzhong Chen are with the School of Remote Sensing and Information Engineering, Wuhan University, Wuhan, Hubei, P.R. China.}% <-this % stops a space
}

\markboth{IEEE TRANSACTIONS ON KNOWLEDGE AND DATA ENGINEERING}%
{Shell \MakeLowercase{\textit{et al.}}: Bare Advanced Demo of IEEEtran.cls for IEEE Computer Society Journals}

\markboth{IEEE TRANSACTIONS ON KNOWLEDGE AND DATA ENGINEERING}%
{Shell \MakeLowercase{\textit{et al.}}: Bare Demo of IEEEtran.cls for Computer Society Journals}

\IEEEtitleabstractindextext{%
\begin{abstract}

Hybrid recommendations have recently attracted a lot of attention where user features are utilized as auxiliary information to address the sparsity problem caused by insufficient user-item interactions in collaborative-filtering-based recommender systems. User features generally contain rich multimodal information, but most of them are irrelevant to the recommendation purpose. Therefore, excessive reliance on these features will make the model overfit on noise and difficult to generalize. In this article, we propose a variational bandwidth auto-encoder (VBAE) for hybrid recommender systems, aiming to address the sparsity and noise problems simultaneously. VBAE first encodes user collaborative and feature information into Gaussian latent variables via deep neural networks to capture non-linear similarities among users. Moreover, by considering the fusion of user collaborative and feature variables as a virtual communication channel from an information-theoretic perspective, we introduce a user-dependent channel to dynamically control the information allowed to be accessed from user feature embeddings. A quantum-inspired uncertainty measurement of hidden rating variables is proposed accordingly to infer the channel bandwidth by disentangling the uncertainty information from the semantic information in user ratings. Through this mechanism, VBAE incorporates adequate auxiliary information from user features if collaborative information is insufficient, while avoiding excessive reliance on noisy user features to improve its generalization ability to new users. Extensive experiments conducted on three real-world datasets demonstrate the effectiveness of the proposed method. Codes and datasets are released at \url{https://github.com/yaochenzhu/VBAE}.

\end{abstract}

\begin{IEEEkeywords}
Recommender systems, uncertainty modeling,
variational inference, information bottleneck, auto-encoders
\end{IEEEkeywords}}

\maketitle

\IEEEdisplaynontitleabstractindextext

\IEEEpeerreviewmaketitle

\IEEEraisesectionheading{\section{Introduction}\label{sec:introduction}}
\label{sec:intro}

\IEEEPARstart{I}n the era of information overload, people have been inundated by large amounts of online content, and it becomes increasingly difficult to discover interesting information. Consequently, recommender systems play a pivotal role in modern online services due to their ability to suggest items that users may be interested in from a large collection of candidates. Based on how recommendations are made, existing recommender systems can be categorized into three classes \cite{zhang2019deep}: collaborative-filtering-based methods, content-based methods, and hybrid methods. Collaborative-filtering-based methods \cite{wang2014relational,chen2020revisiting} predict user preferences by exploiting their past activities, such as browses, clicks, or ratings, where recommendation quality relies heavily on peer users with similar behavior patterns. Content-based methods \cite{xu2013emr, yi2021cross, dong2021dual}, on the other hand, make recommendations based on users or items that share similar features. Hybrid methods \cite{sun2018attentive,xie2020multimodal,chen2020learning,xu2021multi} combine the advantages of both worlds where the collaborative information and user/item features are comprehensively considered to generate more accurate recommendations.

Recent years have witnessed an upsurge of interest in employing auto-encoders \cite{sedhain2015autorec} to both collaborative and content-based recommender systems, where compact representations of sparse user ratings \cite{liang2018variational, li2015deep} or high-dimensional user/item features \cite{wang2016collaborative, zhu2019improving} can be learned to more effectively exploit the similarity patterns between users or items for recommendations. As a Bayesian version of auto-encoder, variational auto-encoder (VAE) \cite{kingma2014auto} demonstrates superiority compared to other forms of auto-encoders, such as contractive auto-encoder \cite{zhang2017autosvd} and denoising auto-encoder \cite{wu2016collaborative}, because they explicitly consider the inferential uncertainty by modeling the latent representations as random variables. Among them,  collaborative variational auto-encoder (CVAE) \cite{li2017collaborative}, which was built upon the collaborative deep learning (CDL) framework \cite{wang2015collaborative}, first used a VAE to infer latent item content embeddings from item textual features. The content embeddings are then finetuned with collaborative information via matrix factorization. Multi-VAE \cite{liang2018variational}, in contrast, used VAE in a collaborative setting to learn compact user embeddings from discrete user ratings. Recently, Macrid-VAE \cite{ma2019learning} further extended Multi-VAE, where learned user representations were constrained to disentangle at both the macro and micro levels to improve the embeddings' interpretability.

\begin{figure}
\centering
\includegraphics[width=0.94\linewidth]{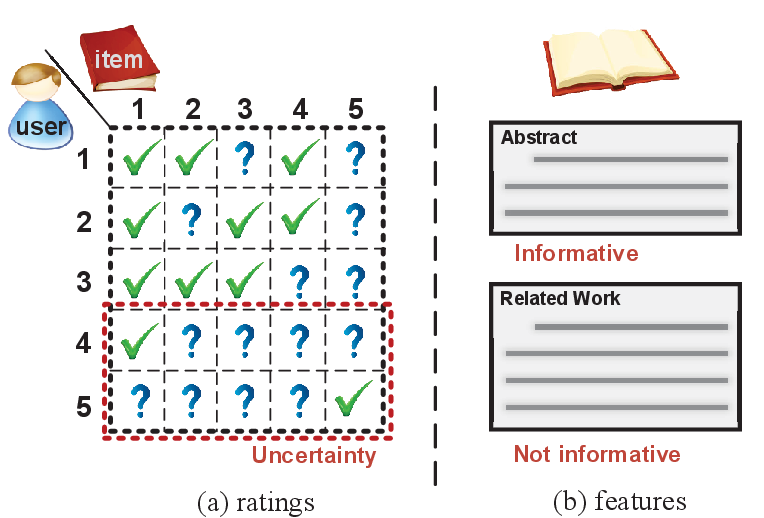}
\caption{Challenges associated with collaborative and feature components of a hybrid recommendation system. Left: Users whose ratings are sparse or who rate only rarely-rated items may have large uncertainty in their collaborative embeddings. Right: User/item features contain large amounts of information that are irrelevant to recommendations.}
\label{fig:teaser}
\end{figure}

Despite the success of VAEs in handling either user rating or feature information, it is difficult to generalize them to a hybrid recommender system, due to various challenges from both the collaborative and content components (Fig. \ref{fig:teaser}). As users tend to vary in their activity levels and tastes, user embeddings learned through collaborative filtering bear different degrees of uncertainty, which hinders good recommendations for users with unreliable collaborative embeddings (\textit{e.g.,} user \#4 and \#5 in Fig. \ref{fig:teaser}). The uncertainty mainly comes from three aspects: (1) \textbf{Sparsity}. For a user with sparser interactions (user \#4), the associated embedding is more unreliable due to the information insufficiency in her historical interactions, which makes the similarity measure induced by the embedding space less informative compared to users with denser interactions. (2) \textbf{Diversity}. Even if a user has denser interactions, we cannot safely conclude that we can estimate her preferences with more confidence, because her ratings may focus on a few types of items, which makes the collaborative information conveyed by different items correlated. In contrast, if she rated items with more diversity, we can have more confidence in estimating her preferences because the item space is more thoroughly explored for the user. (3) \textbf{Overlapping}. In addition, the uncertainty of user embeddings may also be large if items that the user has interacted with are seldomly visited by other users (user \#5). Considering two users who clicked the same number of items, the items clicked by the former user were also interacted with by many other users, while the latter user only clicked items that no other users had yet clicked, the embedding uncertainty of the latter user would be larger than that of the former user. 

Although user/item features could be exploited to reduce the uncertainty incurred by sparse interactions, the main obstacle for utilizing them lies in the heavy noise that may outweigh the useful information. Here, by noise, we mean any pattern that is irrelevant to the recommendation purpose, which should be distinguished from low-level noise such as image blur, audio aliasing, and textual typos. Consider, for example, recommending academic articles to researchers. The related work and empirical study sections of an article are less informative compared to its abstract and methodology, and generally, they should be regarded as noise in recommendations, although they both contain valuable information once a researcher gets attracted by the abstract and decides to delve deeper into the article. Moreover, since different users may consider disparate aspects when rating an item, such noise exhibits a personalized characteristic that makes it extremely difficult to eliminate \cite{hou2019explainable, yi2021dual}. A similar analysis could be made for user features since the collection of certain user attributes, such as location, \textit{etc.,} may raise up privacy issues, and a widely adopted surrogate strategy is to empirically combine the features of items that users have interacted with to build up the user profiles for recommendations. In fact, a consensus in the community is that collaborative filtering is more reliable than feature-based methods for large-scale web recommendations if user interactions are sufficient to leverage \cite{slaney2011web}. Therefore, a good hybrid recommender system should manage to avoid unnecessary reliance on noisy user/item features based on the sufficiency level of collaborative information, such that the noise in these features would not outweigh the useful information and degenerate model generalization ability. 

To address the above challenges, we propose an information-driven Bayesian generative model called variational bandwidth auto-encoder (VBAE)  for hybrid recommender systems. The model first learns the generation of user ratings and features from the corresponding latent Gaussian embeddings via deep neural networks through the auto-encoding variational Bayes (AEVB) algorithm \cite{kingma2014auto}. Furthermore, observing that extracted user features generally contain lots of irrelevant information and could be extremely noisy, we consider the fusion of user collaborative and feature embeddings from an information-theoretic perspective, \textit{i.e.,} as a virtual communication channel. Specifically, the channel is designed to be user-dependent in that it dynamically controls the amount of information allowed to be accessed from user features based on the collaborative information already contained in user interactions. A quantum-inspired uncertainty measurement of hidden rating embeddings is proposed accordingly to infer the channel's bandwidth by disentangling the uncertainty information from the semantic information in historical user ratings. Through this mechanism, sufficient auxiliary information can be accessed from user features when collaborative information is inadequate, while unnecessary dependence of the model on noisy user features can be avoided to improve its generalization to user features. The main contribution of this article can be summarized as follows:

\begin{itemize}
    \item We present VBAE, an information-driven hybrid recommendation framework where the generation of user ratings and features is parameterized via deep Bayesian networks and their fusion is modeled as a user-dependent virtual communication channel, such that the rating sparsity and feature noise problems can be addressed simultaneously.
    \item A quantum-inspired uncertainty measurement of hidden rating embeddings is proposed accordingly to infer the bandwidth of the user-dependent channel, which enhances the model generalization ability by dynamically controlling the information allowed to be accessed from user features based on the sufficiency level of collaborative information.
    \item Two channel implementation strategies with different desired properties, \textit{i.e.,} Bernoulli and Beta channels, are thoroughly discussed, with the corresponding optimization objectives derived with posterior approximation and variational inference to make them amenable to stochastic gradient descent. 
    \item The proposed VBAE empirically out-performs state-of-the-art hybrid recommendation baselines. We also discover that the inferred bandwidth of the channel variable can well distinguish users with different sufficiency levels of collaborative information.
\end{itemize}

\section{Related Work}
\label{sec:rel}

As a special kind of deep neural network, auto-encoders aim to learn compact representations of data by unsupervised self-reconstruction \cite{baldi2012autoencoders}. Since both ratings and user features tend to be high-dimensional sparse vectors, it is difficult to directly manipulate them in the original space, and considerable effort has been dedicated to using auto-encoders to learn their low-dimensional compact representations \cite{sedhain2015autorec}. Generally, based on whether the auto-encoder is used to tackle user or item side information, existing auto-encoder-based recommender systems can be categorized into two main classes: user-oriented auto-encoders (UAEs) \cite{lee2017augmented,liang2018variational,sachdeva2019sequential} and item-oriented auto-encoders (IAEs) \cite{sedhain2015autorec,zhang2020content}. 

\subsection{Item-oriented Auto-encoders}

The advent of IAE predates that of UAE, where item content auto-encoders are built on top of matrix factorization (MF)-based collaborative-filtering backbones, such as weighted matrix factorization \cite{mnih2008prob}, to incorporate auxiliary item content information into factorized item collaborative embeddings. Two exemplar methods from this category are CDL \cite{wang2015collaborative} and CVAE \cite{li2017collaborative}, where an item offset variable is introduced to tightly couple Bayesian stacked denoising auto-encoder (SDAE) \cite{vincent2010stacked} or variational auto-encoder (VAE) \cite{kingma2014auto} with MF to enhance its performance. MF and the item content auto-encoder are then trained in an iterative manner. Recently, auto-encoders have also been exploited to model the item collaborative information. Among them, DICER \cite{zhang2020content} was proposed to capture non-linear item similarity based on their user ratings, where item content information can be disentangled for better generalization. Since for collaborative IAEs, the input dimension and the number of trainable weights are proportional to the number of users while the number of training samples equals the number of items, these methods require a large item-to-user ratio to learn good item representations for recommendations. 

\subsection{User-oriented Auto-encoders}

Compared with IAEs, UAEs have attracted more attention among researchers because they break the long-standing linear collaborative modeling bottleneck of MF-based recommender systems and allow modeling users in a deeper manner \cite{lee2017augmented, wu2016collaborative}. Instead of linearly factorizing the rating matrix into the user and item parts, UAE-based recommenders take the historical ratings of users as inputs, embed them into hidden user representations with a deep encoder network, and reconstruct the ratings with a deep decoder network. The reconstructed ratings for unrated items are then ranked for recommendations. Since UAE-based recommenders eliminate the need of modeling item latent variables and predict the whole ratings directly from user latent embeddings, another advantage of UAEs over MF is that they are efficient to fold in new users with recorded historical ratings because recommendations can be made with a single forward propagation. The first UAE-based recommender system is the collaborative denoise auto-encoder (CDAE) \cite{wu2016collaborative}, where the input rating vectors are randomly masked with zeros before reconstruction to simulate the missing ratings. Afterward, Multi-VAE \cite{liang2018variational} was proposed where a VAE with multinomial likelihood is designed to replace the DAE in CDAE, which leads to improved performance. However, one key problem for all collaborative UAEs is that if ratings of certain users are sparse, the recommendation performance could be severely degenerated due to the lack of collaborative information. 

\subsection{Hybrid Recommendation Techniques}

Due to the wide availability of user features and various methods to build user profiles from features of items that users have interacted with, incorporating auxiliary user feature information into UAE to address the sparsity problem has become a new trend. A simple method is the early fusion strategy \cite{koren2008factorization}, where user features are concatenated with ratings as inputs to UAE. With this strategy, the first dense layer of the UAE can be viewed as fusing user ratings and features via weighted combination, where the weights for the two modalities are trainable and identical for all users. A more sophisticated approach is the conditional VAE (CondVAE) \cite{pang2019novel}, where user features are exploited to calculate the conditional prior of user latent variables. However, both methods treat the relative importance of collaborative and content information as fixed for all users, which ignores individual differences in the sufficiency level of collaborative information and the general low reliability of user features due to pervasive noise \cite{yi2021dual}. 

To address the above two problems, a good and flexible recommender system should avoid unnecessary dependence on user features when collaborative information is sufficient to reduce overfitting on feature noise and improve generalization. Recently, attention mechanism has been introduced into recommender systems to dynamically fuse user ratings and features via user-specific attention weights \cite{li2020adversarial, chen2020learning}. However, the attention weights are generally calculated by modality embedding and softmax normalization, which may fail to capture the relative importance between the rating and feature modalities. Therefore, it motivates us to design a quantum-inspired uncertainty measurement of collaborative information and a channel with uncertainty-related user-specific bandwidth to dynamically regulate the fusion of user feature and collaborative information.

\section{Methodology} 
\label{sec:meth}

\begin{figure*}
\centering
\includegraphics[width=0.85\linewidth]{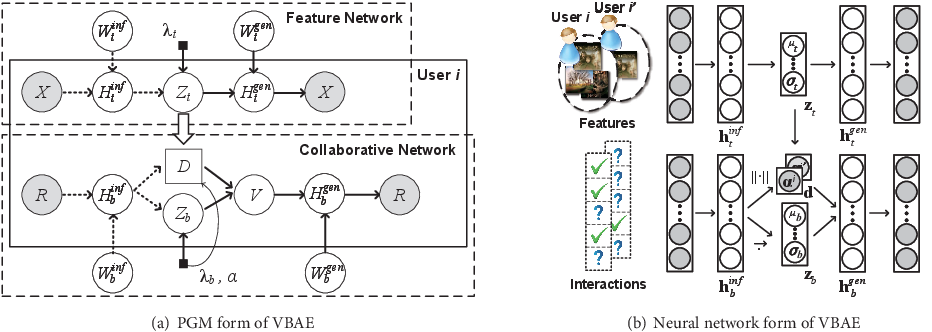}
\caption{Left: The PGM of the proposed VBAE. Right: The zoomed-in view of the collaborative and user feature networks. User $i$ has sparser ratings than user $i'$, which leads to a larger uncertainty in her inferred collaborative embedding, and VBAE infers a larger bandwidth to allow more information of $\mathbf{z}_t$ to flow into $\mathbf{v}$. For user $i'$, on the other hand, a smaller bandwidth is inferred to avoid overfitting on feature noise.}
\label{fig:vbae_pgm}
\end{figure*}

\subsection{Problem Formulation}

The focus of this article is on recommendations with implicit feedback \cite{hu2008collaborative}. We define the rating matrix as $\mathbf{R} \in \mathbb{R} ^ {I \times J}$, where each row vector $\mathbf{r} _ {i}^{T}$ is the bag-of-words vector denoting whether user $i$ visited each of the $J$ items. $\mathbf{R}$ is obtained by keeping track of user activities for a certain amount of time. In addition, the user profiles are represented by matrix $\mathbf{X} \in \mathbb{R} ^ {I \times S}$, where each row vector $\mathbf{x}_{i}^{T}$ is the extracted features for the $i$th user. $\mathbf{x}_{i}$ may include inherent user attributes such as age, location, and self-description, \textit{etc.,} or may be built from the feature of items that the user has interacted with when such information is not available. The capital non-boldface letters $R_{i}$ and $X_{i}$ are used to denote the corresponding random vectors, respectively\footnote{The subscript $i$ will be omitted for simplicity if no ambiguity exists.}. The density of 
$\mathbf{r}_{i}$, which is defined as $\sum_{j}r_{ij}$, could vary dramatically for different $i$, and $\mathbf{x}_{i}$ is generally high-dimensional and noisy. Given the partial observation of $\mathbf{R}$ and user features $\mathbf{X}$, this article focuses on predicting the remaining ratings in $\mathbf{R}$ such that new relevant items can be recommended to users.

\subsection{Model Overview} \label{sec:vbae}

The probabilistic graphical model (PGM) in the left part of Fig. \ref{fig:vbae_pgm} shows the overall generative and inference process of VBAE. The detail is discussed in the following sections.

\subsection{Generative Process} \label{sec:generative}

\subsubsection{Fusion via Channel with User-dependent Bandwidth}

In this article, user collaborative embedding $\mathbf{z}_{b}$ and user feature embedding $\mathbf{z}_{t}$ are assumed to lie in $K$-dimensional Gaussian latent spaces. Traditional generative models for hybrid recommendations \cite{wang2015collaborative, li2017collaborative, dong2017hybrid} directly add $\mathbf{z}_{t}$ and $\mathbf{z}_{b}$ via an offset variable to form the latent user embedding $\mathbf{v}$, or learn their fixed relative weights by concatenation. These methods do not consider that the uncertainty of $\mathbf{z}_{b}$ could vary drastically for different users due to both explicit (rating sparsity) and implicit (rating diversity or overlapping) reasons, nor that $\mathbf{z}_{t}$ contains lots of irrelevant information that may distract the recommendation model. This could be problematic, since for users who have denser ratings or who mainly rate items that are also rated by many other users, their associated collaborative embeddings $\mathbf{z}_{b}$ are more reliable, and $\mathbf{v}$ can afford to access less information from $\mathbf{z}_{t}$ such that unnecessary dependence of the model on noisy user features can be avoided. 

\begin{figure}
\centering
\includegraphics[width=0.94\linewidth]{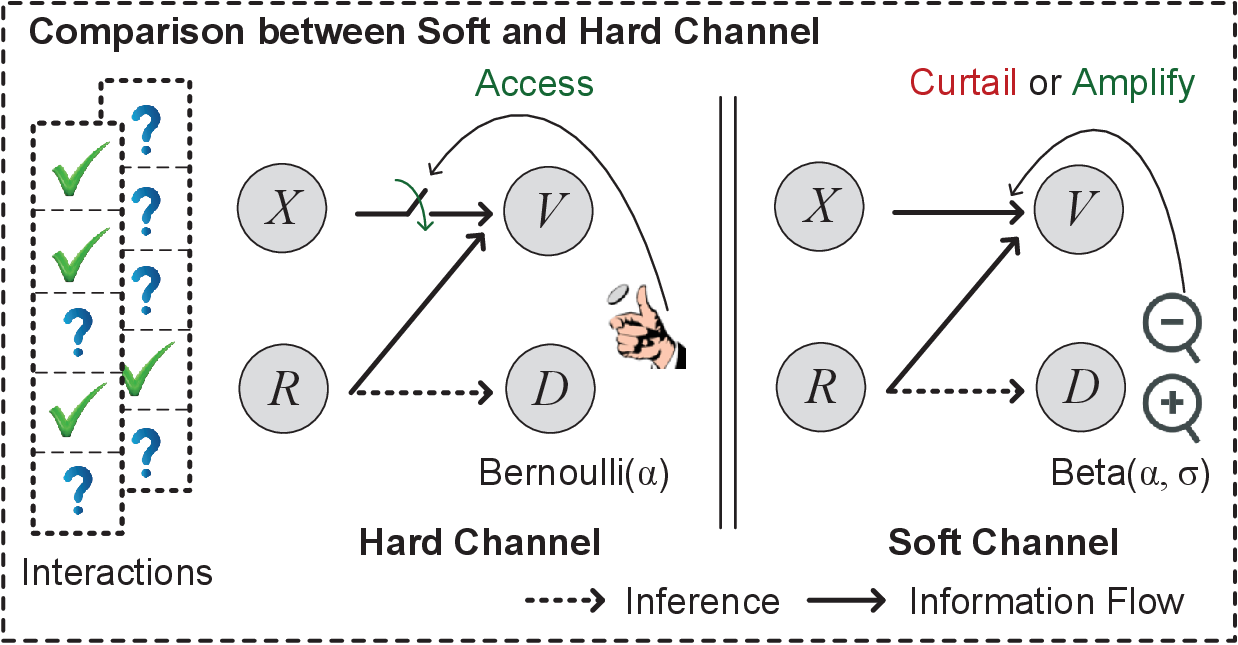}
\caption{Comparisons between the hard and soft channels, where the former conforms more strictly to the variational information bottleneck theory while the latter resembles more to the variational attention mechanism.}
\label{fig:channel}
\end{figure}

From an information-theoretic perspective, if we view the fusion of $\mathbf{z}_{t}$ and $\mathbf{z}_{b}$ into $\mathbf{v}$ as a virtual communication channel, the issue sources from the assumption that the channel is deterministic and independent of the information already contained in the collaborative embedding $\mathbf{z}_{b}$, where the individual difference in the sufficiency level of collaborative information is ignored. To address such a problem, we design a user-dependent channel in VBAE by introducing a latent capacity variable $\alpha$ that determines the channel's bandwidth from $\mathbf{z}_{t}$ to $\mathbf{v}$. The bandwidth $\alpha$ encodes for each user the model's belief towards how much extra information is required from user features given the information already contained in observed interactions. Through this mechanism, the channel dynamically determines the amount of information allowed to flow from $\mathbf{z}_{t}$ to $\mathbf{v}$ conditional on $\mathbf{z}_{b}$. Two strategies are explored to implement the channel. The first strategy is the "hard" channel, where the bandwidth is achieved when $\mathbf{v}$ losslessly accesses $\mathbf{z}_{t}$ with probability $\alpha$ and accesses no information otherwise \cite{Goyal2020The}. In the generative case, we can introduce an auxiliary channel variable $d$ and draw $d$ from the following Bernoulli distribution:
\begin{equation}
\nonumber
    d \sim Bernoulli(\alpha).
\end{equation}
\noindent Although the hard channel conforms more strictly to the definition of bandwidth in information theory, it may result in training instability because the bandwidth $\alpha$ only appears as a statistical property, \textit{i.e.,} an expectation when user feature and collaborative embeddings are repeatedly fused for multiple times. However, for one user, $\mathbf{v}$ either accesses the feature information or not in one specific generation step, which is coarse in granularity to distinguish users with different uncertainty levels of collaborative information. Therefore, we consider a second strategy, namely, the "soft" channel, which is a relaxed version of the hard channel and resembles more to the variational attention proposed in \cite{deng2018latent} than the variational information bandwidth theory \cite{Goyal2020The} that inspires us to design VBAE. This strategy assumes that the channel variable $d$ is drawn from a Beta distribution and the bandwidth $\alpha$ determines the mean of the Beta,
\begin{equation}
\nonumber
    d \sim Beta(\alpha_{1}, \alpha_{2}), \  \text{where} \  \alpha = \alpha_{1} / (\alpha_{1}+\alpha_{2}),
\end{equation}
\noindent and the channel $d$ curtails or amplifies the weights of user feature information based on the bandwidth $\alpha$. Given only $\alpha$, however, the Beta distribution for the channel $d$ is undetermined, as its variance remains to be specified to calculate both $\alpha_{1}$ and $\alpha_{2}$. However, since we care primarily about the bandwidth itself than its uncertainty, we fix the variance of the Beta, which is treated as a nuisance parameter, to a small value $\delta^{2}_{fixed}$ for simplicity). Hereafter, we use the mean-variance parameterization of Beta distribution unless otherwise specified, since it explicitly contains the bandwidth as its first parameter. We use VBAE-hard and VBAE-soft to distinguish the two channel implementation strategies. The detailed comparisons between the soft and hard channels are summarized in Fig. \ref{fig:channel}. After drawing $d$, the user latent variable $\mathbf{v}$ is deterministically calculated as
\begin{equation}
\nonumber
\mathbf{v}=\mathbf{z}_{b} + d \cdot \mathbf{z}_{t},    
\end{equation}
\noindent which defines the fusion process of $\mathbf{z}_{t}$ into $\mathbf{z}_{b}$ via the user-dependent channel of VBAE-hard and VBAE-soft. 

\subsubsection{Neural Network Implementation}

To model the non-linear generation process of user ratings and features from the corresponding latent Gaussian variables, we parameterize the generative distributions as deep neural networks. The user feature $\mathbf{x}$ is generated from the user feature latent variable $\mathbf{z}_{t}$ via a multilayer perceptron (MLP) $\operatorname{Gen} _ {t} (\mathbf{z}_{t})$:  If $\mathbf{x}$ is binary, we squashed the outputs of $\operatorname{Gen} _ {t}$ by the sigmoid function, and draw $\mathbf{x}$ from $Bernoulli(\operatorname{sigmoid}(\operatorname{Gen} _ {t} (\mathbf{z}_{t})))$, or if $\mathbf{x}$ is real-value, we take the raw outputs of $\operatorname{Gen} _ {t}$ as the mean of a Gaussian distribution and draw $\mathbf{x}$ from $\mathcal{N}\left(\operatorname{Gen} _ {t} (\mathbf{z}_{t}), \lambda_{x}^{-1} \mathbf{I}_{S}\right)$. Finally, we put a multinomial likelihood on ratings $\mathbf{r}$ as \cite{liang2018variational} and generate $\mathbf{r}$ from the user latent variable $\mathbf{v}$ via $p(\mathbf{r}|\mathbf{v})$ parameterized as $Multi(N_{int}, \operatorname{softmax}(\operatorname{Gen}_{b}(\mathbf{v})))$, where $\operatorname{Gen}_{b}$ is another MLP-based generative neural network, and $N_{int}$ is the number of interacted items. The generation process of $\mathbf{x}$, $\mathbf{r}$ from $\mathbf{z}_{t}$, $\mathbf{v}$ is provided as follows:

(1) \ For each layer $l$ of the collaborative and the user feature modules of the generation network:

\quad \quad  (a) \ For each column $n$ of the weight matrices, draw
\begin{equation}
\nonumber
    \mathbf{W}^{(l)}_{\{b,t\}, : n} \sim \mathcal{N}\left(\mathbf{0}, \lambda_{w}^{-1} \mathbf{I}_{K_{l}}\right);
\end{equation}
\quad \quad  (b) \ Draw the bias vector from $\mathbf{b}^{(l)}_{\{b, t\}} \sim \mathcal{N}\left(\mathbf{0}, \lambda_{w}^{-1} \mathbf{I}_{K_{l}}\right)$;
\quad \quad  (c) \ For ${\mathbf{h}}^{(l)}_{\{b,t\}}$ of a user $u$, draw
\begin{equation}
\nonumber
  \label{EQ:INTER}
  \mathbf{h}^{(l)}_{\{b,t\}} \sim \delta\left(\operatorname{f_{act}}\left(\mathbf{W}^{(l)}_{\{b,t\}} \mathbf{h}^{(l-1)}_{\{b,t\}}  +\mathbf{b}^{(l)}_{\{b, t\}}\right)\right).
\end{equation}
(2) For user features that are binary, draw
\begin{equation}
\nonumber
    \mathbf{x} \sim Bernoulli\left(\operatorname{sigmoid}\left(\mathbf{W}^{(L+1)}_{t} \mathbf{h}_{t}^{(L)} +\mathbf{b}^{(L+1)}_{t}\right)\right);
\end{equation}
\quad \  For user-item interactions, draw
\begin{equation}
\nonumber
    \mathbf{r} \sim Multi\left(N_{int}, \operatorname{softmax}\left(\mathbf{W}^{(L+1)}_{b} \mathbf{h}_{b}^{(L)} +\mathbf{b}^{(L+1)}_{b}\right)\right),
\end{equation}
\noindent where $\mathbf{h}_{t}^{(0)}=\mathbf{z}_{t}$, $\mathbf{h}_{b}^{(0)}=\mathbf{v}$, $\lambda_{w}$ is a hyperparameter, $\mathrm{f}_{act}$ is the intermediate activation function and $\delta$ is the Dirac Delta function. Step 1.c can be alternatively viewed as putting Gaussian priors on the intermediate activations and setting the precision to infinity.

The generative model of VBAE is described by the joint distribution of all observed and hidden variables,
\begin{equation}
\label{eq:gen}
\begin{aligned}
    p_{\theta}(R, X, Z_{b} , Z_{t} , D) = & p _{\theta} \left(R|Z_{b} , Z_{t} , D\right) p_{\theta}\left(X | Z_{t}\right)\\ &p\left( Z_{b}\right)p\left(Z_{t}\right)p\left( D\right),
\end{aligned}
\end{equation}

\noindent where the symbol $\theta$ denotes the set of trainable weights that pertain to the generative network.

\subsection{Inferential Process} \label{sec:infer} Given Eq. (\ref{eq:gen}), however, it is intractable to calculate the posterior $p_{\theta}(Z_{b} , Z_{t} , D|R, X)$, as the non-linearity of generative process precludes us from integrating over the latent space and calculating the marginal $p_{\theta}(R, X)$. Therefore, we resort to amortized variational inference \cite{blei2017variational}, where we introduce a variational posterior $q_{\phi}(Z_{b} , Z_{t} , D|R, X)$ parameterized by an inference network with trainable weights $\phi$ as an approximation to the true but intractable posterior. According to the conditional independence assumptions implied by VBAE, the joint variational posterior can be decomposed into the compact product of two factors that make up two modules of the inference network: the collaborative module $q_{\phi}(Z_{b}, D|R)=q_{\phi}(Z_{b}|R) \times q_{\phi}(D|R)$ that infers the user collaborative embeddings and the user-dependent channel variables from user ratings, and the user feature module $q_{\phi}(Z_{t}|X)$ that infers user feature embeddings from user features, respectively. 

\subsubsection{Quantum-Inspired Semantic and Uncertainty Measurement of Collaborative Information}

\begin{figure}
\centering
\includegraphics[width=0.93\linewidth]{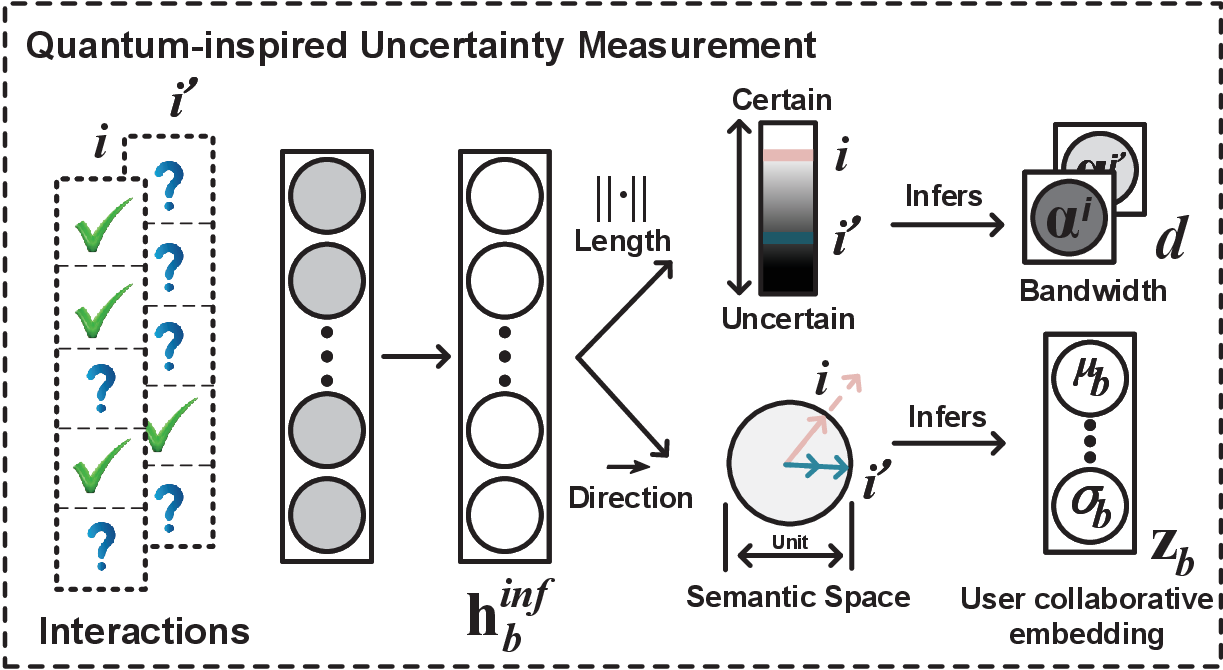}
\caption{Illustration of the proposed quantum-inspired semantic and uncertainty measurement of collaborative information from observed user interactions.}
\label{fig:uncertainty}
\end{figure}

The collaborative module of VBAE infers the latent collaborative and channel variable from the observed user-item interactions. Since the channel's bandwidth corresponds to the insufficiency level of collaborative information, an important function of this module is to disentangle the uncertainty and semantic information in user ratings. To achieve this objective, we first use an MLP to embed the raw rating vector $\mathbf{r}$ into a compact hidden representation $\mathbf{h}^{inf}_{b}$ (the superscript $inf$ that denotes the inference process will be omitted for simplicity if no ambiguity exists) for each user. Inspired by \cite{li2019cnm}, we use the length (L2-norm) of $\mathbf{h}_{b}$ as the uncertainty measurement of collaborative information to calculate the bandwidth and use the direction of $\mathbf{h}_{b}$ (\textit{i.e.,} L2-normalized $\mathbf{h}_{b}$) as the representation of semantic information to infer the latent collaborative embedding. The detail of the introduced semantic and uncertainty measurement of collaborative information is illustrated in Fig. \ref{fig:uncertainty}, which draws inspiration from theoretic quantum mechanics.

To see the connections, we first assume a mixed-state physical system can be represented by the identity matrix $\mathbf{I}_{K_{h}}$ and liken the hidden rating embedding $\mathbf{h}_{b}$ to a quantum superposition state of the  system, which is represented by a complex vector in $\mathbb{C}^{K_{h}}$. In such a case, the superposition vector has the property that its norm is positively correlated with the probability of observing this superposition state when measuring the system (which depicts the fundamental uncertainty of quantum physics), and its direction distinguishes the state from other superposition states (which carries the state's semantic information).

Such a semantic-uncertainty interpretation of a vector's length and direction also works for hidden rating embedding $\mathbf{h}_{b}$ in recommendations because the length and direction of $\mathbf{h}_{b}$ can be associated with similar meanings. To gain the intuition, we analyze the case where the network used to calculate $\mathbf{h}_{b}$ has one linear dense layer (Generalization to the multi-layer case requires linearizing the network locally with network Jacobians). We first decompose the calculation of $\mathbf{h}_{b}$ from raw rating vector $\mathbf{r}$ into two basic operations, embedding and element-wise sum, as follows:
\begin{equation}
    \mathbf{h}_{b} = \mathbf{W}^{(1)}_{b} \cdot \mathbf{r} = \sum_{j:r_{j}=1}\mathbf{w}^{(1)}_{b,:j}.
\end{equation}
\noindent If we assume that each element in $\mathbf{W}^{(1)}_{b}$, \textit{i.e.,} $w^{(1)}_{b,ij}$ is independent and identically distributed (i.i.d.) Gaussian variable with zero mean and a small variance $\epsilon^{2}$, each element in $\mathbf{h}_{b}$, \textit{i.e.,} $h_{b,k}$, is the sum of $N_{int}$ independent Gaussian variables, where $N_{int} = \sum_{j:r_{j}=1} 1$ is the number of items the user has interacted with, \textit{i.e.,} the rating density. Therefore, $h_{b,k}$ is also a Gaussian variable that follows $\mathcal{N}(0, N_{int} \epsilon^{2})$. According to the probability theory, since the squared L2-norm of $\mathbf{h}_{b}$, \textit{i.e.,} $\| \mathbf{h}_{b} \|^{2}$, is the sum of squares of $K_{h}$ Gaussian variables with zero mean and $N_{int} \epsilon^{2}$ variance, $\| \mathbf{h}_{b} \|^{2}$ follows the scaled Chi-square distribution denoted as $N_{int} \epsilon^{2} \cdot \chi_{K_{h}}^{2}$, which is equivalent to Gamma distribution with cumulative distribution function $\Gamma(K_{h} / 2,2 N_{int} \epsilon^{2})$ \cite{Johnson2003}. Finally, according to the property of Gamma distribution, the expected value of $\| \mathbf{h}_{b} \|^{2}$ can be calculated as $N_{int} K_{h} \epsilon^{2}$. This term is a monotonic increasing function of the number of interacted item $N_{int}$, \textit{i.e.,} the rating density, which is a main indicator for the sufficiency level of collaborative information. 

\subsubsection{Bandwidth and User Collaborative Embedding}

The above property reveals that $\| \mathbf{h}_{b} \|^{2}$ is positively correlated with the sufficiency level of collaborative information. Moreover, compared to the sparse rating vector $\mathbf{r}$, the hidden embedding $\mathbf{h}_{b}$ lies in a compact, low dimensional space where redundant information can be eliminated. In addition, the L2-normalization of $\mathbf{h}_{b}$ eliminates the negative influence of individual differences in interaction density by scaling the hidden rating embeddings of users with different activity levels into the same sphere space. This makes the direction of $\mathbf{h}_{b}$ a more suitable representation than the original $\mathbf{h}_{b}$ to infer user collaborative embeddings. This is in contrast with previous auto-encoder-based approaches such as Multi-VAE \cite{liang2018variational} and CondVAE \cite{pang2019novel}, where the information regarding the sparsity level of the historical ratings is discarded after the L2-normalization of input ratings. 

In implementation, we calculate the bandwidth by a linear transformation of $\| \mathbf{h}_{b} \|$ with sigmoid function to squash the value between [0, 1],
\begin{equation}
\label{eq:bd_inf}
    \alpha = \operatorname{sigmoid} \left( w^{\alpha}_{b} \| \mathbf{h}_{b} \| + b^{\alpha}_{b}\right).
\end{equation}
\noindent Since $\| \mathbf{h}_{b} \|$ is strictly non-negative, in backward propagation, the weight $w^{\alpha}_{b}$ can only be updated in one direction if the sign of gradients is the same for the users in a mini-batch. This is problematic, since the training loop could make $\alpha$ converge to the extremity (\textit{i.e.,} zero or one) where the bandwidth for all users is identical, which fails to distinguish users with different sufficiency levels of collaborative information. This is referred to as the mode collapsing problem in variational inference. In this article, this problem is addressed by batch normalization \cite{ioffe2015batch}. We denote a mini-batch of length of user hidden embeddings as $\mathcal{B} = \{\| \mathbf{h}_{b, 0} \|, \| \mathbf{h}_{b, I_{mini}-1} \|, \cdots, \| \mathbf{h}_{b, I_{mini}} \|\}$. Before calculating the bandwidth from samples in $\mathcal{B}$ as Eq. (\ref{eq:bd_inf}), we renormalize them by
\begin{equation}
    \widehat{\| \mathbf{h}_{b, i}\|} \leftarrow \frac{\| \mathbf{h}_{b, i}\|-\mu_{\mathcal{B}}}{\sqrt{\sigma_{\mathcal{B}}^{2}+\epsilon}},
\end{equation}
\noindent where $\mu_{\mathcal{B}}$ and $\sigma_{\mathcal{B}}^{2}$ are the mean and sample variance of $\| \mathbf{h}_{b}\|$ in the mini-batch, and $\epsilon$ is a small value to avoid the zero-division error. By batch normalization, $\mathcal{B}$ is balanced in positive and negative samples, which leads to more stable training dynamics and improves the consistency between the inferred bandwidth $\alpha$ and the sufficiency level of collaborative information. In the testing phase, $\mu_{\mathcal{B}}$ and $\sigma_{\mathcal{B}}^{2}$ are fixed to their estimated running average.

\subsubsection{Neural Network Implementation}

The user feature module infers latent feature embeddings from the extracted user features by another MLP, which serves as the auxiliary information source to the collaborative information. The detailed inference process of $Z_{b}, Z_{t}$, and $D$ is described as follows:

(1) \ For each layer $l$ of the collaborative and user feature module of the inference network:

\quad \quad  (a) \ For each column $n$ of the weight matrices, draw
\begin{equation}
\nonumber
    \mathbf{W}^{(l)}_{\{b,t\}, : n} \sim \mathcal{N}\left(\mathbf{0}, \lambda_{w}^{-1} \mathbf{I}_{K_{l}}\right);
\end{equation}
\quad \quad  (b) \ Draw the bias vector from $\mathbf{b}^{(l)}_{\{b, t\}} \sim \mathcal{N}\left(\mathbf{0}, \lambda_{w}^{-1} \mathbf{I}_{K_{l}}\right)$;

\quad \quad  (c) \ For ${\mathbf{h}}^{(l)}_{\{b,t\}}$ of a user $i$, draw
\begin{equation}
\nonumber
  \mathbf{h}^{(l)}_{\{b,t\}} \sim \delta\left(\operatorname{f_{act}}\left(\mathbf{W}^{(l)}_{\{b,t\}} \mathbf{h}^{(l-1)}_{\{b,t\}}  +\mathbf{b}^{(l)}_{\{b, t\}}\right)\right).
\end{equation}
(2) \ For the user-dependent channel variable:

\quad \quad   (a)  Calculate the bandwidth from its logits inferred by the L2-norm of the hidden rating embedding $\mathbf{h}^{(L)}_{b}$ as:
\begin{equation}
\nonumber
    \alpha = \operatorname{sigmoid} \left(w^{\alpha}_{b} \| \mathbf{h}^{(L)}_{b} \|  + b^{\alpha}_{b}\right);
\end{equation}
\quad \quad   (b)  For VBAE-hard, draw the Bernoulli channel variable as:
\begin{equation}
\nonumber
    d \sim Bernoulli (\alpha);
\end{equation}
\quad \quad   (c)  For VBAE-soft, draw the Beta channel variable as:
\begin{equation}
\nonumber
    d \sim Beta (\alpha, \sigma^{2}_{fixed}).
\end{equation}

(3) \ For the collaborative and user feature latent variable as:

\quad \quad   (a)  Draw the mean and standard deviation as:
\begin{equation}
\nonumber
\left[\boldsymbol{\mu}_{\{b, t\}},\log \boldsymbol{\sigma}_{\{b, t\}}\right]  \sim \delta \left(\mathbf{W}^{\mu+\sigma}_{\{b,t\}} \mathbf{h}^{(L)}_{\{b,t\}}   +  \mathbf{b}^{\mu+\sigma}_{\{b,t\}}\right);
\end{equation}
\quad \quad   (b)  Draw the sample of the latent variable:
\begin{equation}
\nonumber
    \mathbf{z}_{\{b, t\}} \sim \mathcal{N}\left(\boldsymbol{\mu}_{\{b, t\}}, \operatorname{diag}\left(\boldsymbol{\sigma}_{\{b, t\}}^{2}\right)\right).
\end{equation}
It is not trivial to allow gradients to be back-propagated to the weights of the inference network through the stochastic Bernoulli or Beta channel variables, and we defer the discussion of the solution, which is called posterior approximation and reparameterization trick, to Section \ref{sec:mcmc}. The neural network implementation of VBAE is illustrated in the right part of Fig. \ref{fig:vbae_pgm} for reference. 

\subsection{Training Objective} To jointly learn the parameters of the generative network and the inference network, we maximize the evidence lower bound (ELBO), which is an approximation to the marginal log-likelihood of the evidence $p_{\theta}(R, X)$ as follows:
\begin{equation}
\label{eq:vbae_obj}
\begin{aligned}
\mathcal{L}& =\mathbb{E}_{q_{\phi}}[\log p_{\theta}(R, X, Z_{b}, Z_{t}, D)-\log q_{\phi}( Z_{b}, Z_{t}, D|R, X)] \\
&= \mathbb{E}_{q_{\phi}}[\log p_{\theta}(R | Z_{b}, Z_{t}, D) +\log p_{\theta}(X | Z_{t})] \\
&-\mathbb{KL}\left[q_{\phi}(Z_{b}, D | R) \| p(Z_{b}, D)\right] - \mathbb{KL}\left[q_{\phi}(Z_{t} | X) \| p(Z_{t})\right],
\end{aligned}
\end{equation}
\noindent and the value of $\mathcal{L}$, for a fixed $\theta$, achieves the maximum if and only if the discrepancy between the variational approximation $q_{\phi}$ and the true posterior $p_{\theta}$ measured by the Kullback-Leiber (KL) divergence is zero (\textit{i.e.}, iif. $q_{\phi}=p_{\theta}$).

\subsection{Maximum A Posteriori Estimation} \label{sec:map_esti}

The collaborative and user feature module of VBAE can be jointly trained as Eq. (\ref{eq:vbae_obj}). However, it could incur extra computational and memory consumption. In addition, in joint training, VBAE may converge to a sub-optima where it relies solely on one information source for recommendations. To address the problems, we utilize an EM-like optimization approach, where we iteratively consider one of the variational distributions in  ${q_{\phi}(Z_{b}, D | R)}$ and $q_{\phi}(Z_{t} | X)$ and fix the random variables that concern the other to their means or previous estimates. Specifically, for the collaborative part ${q_{\phi}(Z_{b}, D | R)}$, we fix $Z_{t}$ to the estimated mean to calculate the user latent variable $V$, and the objective becomes:
\begin{equation}
\label{eq:bstep}
\begin{aligned}
&\mathcal{L}_{b\_\mathrm{step}}^{MAP} = \mathbb{E}_{q_{\phi}(Z_{b}, D | R)}[\log p(R | V)]  - \mathbb{KL}\left[q_{\phi}( D | R) \| p(D)\right] \\ & -\mathbb{KL}\left[q_{\phi}(Z_{b} | R) \| p(Z_{b})\right] -\frac{\lambda_{w}}{2} \cdot  \sum_{l}\left(\|\mathbf{W}_{b}^{(l)}\|_{F}^{2}+\|\mathbf{b}_{b}^{(l)}\|_{2}^{2}\right),
\end{aligned}
\end{equation}
\noindent where $V = Z_{b} + D \cdot \boldsymbol{\mu}_{t}$, and the prior of $D$ is $Bernoulli(0.5)$ for VBAE-hard and is the standard logistic-normal distribution for VBAE-soft. Eq. (\ref{eq:bstep}) can be viewed alternatively as paying a cost equals to  $\mathbb{KL}\left[q_{\phi}( D | R) \| p(D) \right]$  whenever the model accesses information from noisy user features, where the cost is dynamically decided by the urgency level of introducing extra information based on the sufficiency level of collaborative information. This can prevent the model from depending excessively upon noisy user features. After one-step optimization of $\mathcal{L}_{b\_\mathrm{step}}^{MAP}$, we then fix $Z_{b}$ and $D$ to their estimated values to calculate $V$, and maximize the following objective for the user feature part of VBAE:
\begin{equation}
\label{eq:tstep}
\begin{aligned}
&\mathcal{L}_{t\_\mathrm{step}}^{MAP} = \mathbb{E}_{q_{\phi}(Z_{t} | X)}[\log p(R | V)] + \mathbb{E}_{q_{\phi}(Z_{t} | X)}\left[\log p(X | Z_{t}) \right] \\ &- \mathbb{KL}\left[q_{\phi}(Z_{t} | X) \| p(Z_{t})\right]  
-\frac{\lambda_{w}}{2}\cdot \sum_{l}\left(\|\mathbf{W}_{t}^{(l)}\|_{F}^{2}+\|\mathbf{b}_{t}^{(l)}\|_{2}^{2}\right).
\end{aligned}
\end{equation}
\noindent Intuitively, the objectives for both steps consist of two parts. The first part is the expected log-likelihood term, which encourages the hidden Gaussian embeddings and the latent channel variable to best explain the observed historical ratings and the extracted user features. The second part is the KL-with-prior terms and the weight decay terms, which act as regularizers to prevent over-fitting.
Liang \textit{et al.} \cite{liang2018variational} have shown that the KL regularization for $Z_{b}$ could be too strong, which over-constrains the representation ability of latent collaborative embeddings. As a solution, they introduced a scalar $\beta$ to control the weight of the KL term in Eq. (\ref{eq:bstep}), which has its theoretical foundation in both beta-VAE \cite{BetaVAE} and variational information bottleneck \cite{DVIP,xie2021micro}. We anneal $\beta$ from 0 to 0.2 as \cite{liang2018variational} and have confirmed the effectiveness of KL annealing in our experiments. Under such settings, the model learns to encode more information of $\mathbf{r}$ in $\mathbf{z}_{b}$ in the initial training stages while gradually regularizing $\mathbf{z}_{b}$ by forcing it close to the prior as the training proceeds \cite{BowmanVVDJB16}.
 
\subsection{Monte Carlo Gradient Estimator} \label{sec:mcmc}

In this section, we derive the gradients of $\mathcal{L}_{b\_\mathrm{step}}^{MAP}$ and $\mathcal{L}_{t\_\mathrm{step}}^{MAP}$ w.r.t. the trainable parameters of the generation and inference network to make them amenable to SGD optimization. For Gaussian and Bernoulli distributions, since their KL divergence with prior can be computed analytically, the minimization of the KL terms in Eqs. (\ref{eq:bstep}), (\ref{eq:tstep}) w.r.t. the weights of the inference network $\phi$ can be properly calculated (For Beta variable, however, posterior approximation is required to calculate the KL term, which will be discussed in the following subsections). However, since the gradients of the expected log-likelihood terms, which we denote as $\operatorname{ELM}_{\{b,t\}\_\mathrm{step}}$, need to be back-propagated through stochastic nodes {$Z_{b}$, $Z_{t}$ and $D$}, it precludes us from calculating an analytic solution. Hence, we introduce Monte Carlo methods to form unbiased estimators of the gradients. 

For $\theta$, as the generative distribution is explicit in an expectation form, its gradient can be estimated by generating samples from the encoder distribution, calculating the gradients, and taking the average \cite{RezendeMW14}. For $\phi$ associated with the inference network, however, no direct sampling strategy is viable to estimate the gradients. Therefore, we resort to the reparameterization trick, where we transform the stochastic nodes into differentiable bivariate functions of their parameters and random noise, and estimate the gradients by averaging the gradients sampled from the noise distribution. The detailed solution for three kinds of random variables used in this article is introduced in the following subsections. 

\subsubsection{User Embeddings: Vanilla Reparameterization}

For the latent Gaussian variables $Z_{\{b, t\}}$, with the vanilla reparameterization trick introduced in \cite{kingma2014auto}, their samples can be reparameterized as:
\begin{equation}
\label{eq:rep_val}
\mathbf{z} _ {\{b, t\}} ^ {(l)}  ([\boldsymbol{\mu} _ {\{b, t\}}, \boldsymbol{\sigma}_{\{b,t\}}], \boldsymbol{\epsilon}^{(l)})= \boldsymbol{\mu} _ {\{b, t\}} + \boldsymbol{\epsilon}^{(l)} \odot \boldsymbol{\sigma}_{\{b,t\}}.   
\end{equation}
\noindent where $\boldsymbol{\epsilon} ^{(l)} \sim \mathcal{N}(0, \mathbf{I}_{K})$. Eq. (\ref{eq:rep_val}) could be viewed alternatively as injecting Gaussian noise to user latent  collaborative and feature variables, which is the main denoising mechanism of previous auto-encoder-based recommender systems to address low-level noise in user ratings and features.

\subsubsection{Hard Channel: Gumbel-Softmax Reparameterization}

For the hard channel that follows the Bernoulli distribution, we note that sampling from which is equivalent to sampling a one-hot vector from a two-class Categorical distribution with probability mass $\boldsymbol{\alpha} = [\alpha, 1-\alpha]$ and discarding the second dimension. Therefore, we resort to the Gumbel-softmax trick \cite{jang2017categorical} and reparameterize the samples as:
\begin{equation}
\label{eq:rep_categ}
\begin{aligned}
d^{(l)}(\boldsymbol{\alpha},  \mathbf{g}^{(l)})&=\frac{\exp \left(\left(\log \left(\alpha_{1}\right)+g^{(l)}_{1}\right) / \tau\right)}{ \sum _ {i=1}^{2}\exp \left(\left(\log \left(\alpha_{i}\right)+g^{(l)}_{i}) / \tau\right) \right)} \\
&= \operatorname{sigmoid}\left(\log \left(\frac{\alpha}{1-\alpha}\right) + g^{(l)}_{1} - g^{(l)}_{2}, \tau \right),
\end{aligned}
\end{equation}
\noindent where $ g^{(l)}_{1} \perp g^{(l)}_{2} \sim Gumbel(0, 1)$ and $\tau$ is the temperature of the softmax and the sigmoid. When $\tau$ approaches zero, the samples $d^{(l)}$ are proved to be equivalent to samples drawn from the corresponding Bernoulli distribution with parameter $\alpha$. In practice, $\tau$ is gradually annealed as the training proceeds for a more stable convergence.

\subsubsection{Soft Channel: Logistic-Normal Reparameterization}

Similarly, we draw the Beta channel of VBAE-soft by keeping the first dimension of a sample from the corresponding two-class Dirichlet distribution. However,  there is no consensus regarding how to reparameterize a Dirichlet variable \cite{joo2020dirichlet, Akash2017Auto, deng2018latent}.  In this article, we eschew the commonly used strategies that transform a uniformly distributed vector by the inverse of the Dirichlet cumulative distribution function as the bivariate transformation. In contrast, we derive its reparameterization with logistic-normal posterior approximation instead \cite{Akash2017Auto}. The reason is that, the logistic-normal distribution converts the original concentration parameters $[\alpha_{1}, \alpha_{2}]$ of the Dirichlet (the values that should be predicted by the inference network) to the mean and standard deviation of a Gaussian distribution, such that the convergence to a low-entropy area is smoother. Otherwise, to reach a low variance area of the Dirichlet requires large values of $[\alpha_{1}, \alpha_{2}]$, which is difficult to learn by the inference network and results in unstable training dynamics \cite{deng2018latent}. The relationship between the parameters of logistic-normal and the corresponding Dirichlet is formulated as follows
\begin{equation}
\label{eq:rep_diri}
\begin{aligned}
&\mu_{1} = -\mu_{2} =\frac{1}{2} \bigg(\log \alpha_{1}- \log \alpha_{2} \bigg) \\
&\sigma_{1} = \sigma_{2} = \frac{1}{4} \bigg(\frac{1}{\alpha_{1}} + \frac{1}{\alpha_{2}}\bigg)= \sigma^{\prime}_{fixed}.
\end{aligned}    
\end{equation}
\noindent We fix the logistic-normal to a small value in Eq. (\ref{eq:rep_diri}), as here only the mean of the Beta channel variable (\textit{i.e.}, the bandwidth) is important, and a small value of the variance prevents the Dirichlet distribution from stuck into a low-entropy area. This is also a common trick used in most regression task, where the outputs are assumed to be the mean Gaussian and the variance is deemed as fixed. The sample $d^{(l)}$ from the logistic-normal is drawn according to
\begin{equation}
\label{eq:rep_beta}
\begin{aligned}
d^{(l)}(\boldsymbol{\alpha},  \boldsymbol{\epsilon}^{(l)})&=\frac{\exp \left(\left( \mu_{1}+\sigma_{1} \cdot \epsilon _ {1} ^ {(l)}\right) \right)}{ \sum _ {i=1}^{2}\exp \left(\left(\mu_{i}+\sigma_{i} \cdot \epsilon _ {i} ^ {(l)}\right) \right)} \\
&= \operatorname{sigmoid}\left( 2 \cdot \mu_{1} + \sigma^{\prime}_{fixed} \cdot (\epsilon^{(l)}_{1} - \epsilon^{(l)}_{2}) \right),
\end{aligned}
\end{equation}
\noindent where $\epsilon^{(l)}_{1} \perp \epsilon^{(l)}_{2} \sim \mathcal{N}(0, 1)$. Eq. (\ref{eq:rep_beta}) can be calculated from bandwidth $\alpha$ because $2 \cdot \mu_{1} = log(\frac{\alpha_{1}}{\alpha_{2}}) = log(\frac{\alpha}{1-\alpha})$, where the first equation holds according to Eq. (\ref{eq:rep_diri}), and the second equation holds because the property of Dirichlet Distribution leads to the equality $\alpha=\frac{\alpha_{1}}{\alpha_{1} + \alpha_{2}}$ ). A close look at Eq. (\ref{eq:rep_beta}) shows that it bears great similarity to Eq. (\ref{eq:rep_categ}), since we can view $2 \cdot \mu_{1} = log(\frac{\alpha}{1-\alpha})$ as pseudo logits of the bandwidth, and both Eqs. add and subtract two i.i.d. random variables to the bandwidth logits before squashing it into (0, 1) with the sigmoid function. The major difference is that in Eq. (\ref{eq:rep_categ}), a small temperature of the sigmoid pushes the value of $d^{(l)}$ to the two extremities, \textit{i.e.,} zero or one, such that for each user the channel is either open or closed in one iteration. However, in Eq. (\ref{eq:rep_categ}), the temperature is fixed at one and $d^{(l)}$ can take any value between [0, 1], which avoids sudden swerve of gradient direction in training and smooths the convergence. 
\begin{algorithm}[t]
\DontPrintSemicolon  \KwIn{$\mathcal{D} = \{(\mathbf{r}_{i}, \mathbf{x}_{i}) \}$, a dataset of collected user ratings and features, where $\mathbf{r}_{i} \in \{0, 1\}^{J}$, $i \in \{1, 2, \cdots, I\}$; $\Theta_{t}$, $\Theta_{b}$, the randomly initialized weights of the collaborative and feature sub-networks; $lr$, the learning rate.}
 \While{metrics on validation users improve,}{
  $\mathcal{L}_{t\_\mathrm{step}}^{MAP} = 0$;\;
  \ForAll{$(\sim, \mathbf{x}_{i}) \in \mathcal{D}$,}{
      Infer $\boldsymbol{\mu}_{t}$ and $\boldsymbol{\sigma}_{t}$ and sample $\mathbf{z}_{t} \sim \mathcal{N}(\boldsymbol{\mu}_{t}, \boldsymbol{\sigma}_{t})$;\;
      Calculate $\hat{\mathbf{x}}_{i}$ the reconstruction of $\mathbf{x}_{i}$ from $\mathbf{z}_{t}$;\;
      Calculate $\mathcal{L}_{i, t\_\mathrm{step}}^{MAP}$ from $\hat{\mathbf{x}}_{i}$ and $\mathbf{x}_{i}$ as Eq. (\ref{eq:tstep}).\;
      $\mathcal{L}_{t\_\mathrm{step}}^{MAP} += \mathcal{L}_{i, t\_\mathrm{step}}^{MAP}$;\;
  }
  
  Update $\Theta_{t} = \Theta_{t} - lr \cdot \nabla _ {\Theta_{t}} (-\mathcal{L}_{t\_\mathrm{step}}^{MAP})  / I$
 
  \ForAll{$(\mathbf{r}_{i}, \mathbf{x}_{i}) \in \mathcal{D}$,}{
      Infer $\boldsymbol{\mu}_{b}$ and $\boldsymbol{\sigma}_{b}$ and sample $\mathbf{z}_{b} \sim \mathcal{N}(\boldsymbol{\mu}_{b}, \boldsymbol{\sigma}_{b})$;\;
      Infer $\boldsymbol{\mu}_{t}$ and $\boldsymbol{\sigma}_{t}$ and sample $\mathbf{z}_{t} \sim \mathcal{N}(\boldsymbol{\mu}_{t}, \boldsymbol{\sigma}_{t})$;\;
      Infer $\alpha$ from $\mathbf{r}_{i}$ and\;
      \quad For VBAE-hard,\;
      \quad \quad Sample $d \sim Bernoulli(\alpha)$;\;
      \quad For VBAE-soft,\;
      \quad \quad Sample $d \sim LogisticNormal(\alpha, \sigma^{\prime}_{fixed})$;\;
      Calculate $\mathbf{v} = \mathbf{z}_{b} + d \cdot \mathbf{z}_{t}$;\;
      Calculate $\hat{\mathbf{r}}_{i}$ the multinomial probability of $\mathbf{r}_{i}$;\;
      Calculate $\mathcal{L}_{i, b\_\mathrm{step}}^{MAP}$ from $\hat{\mathbf{r}}_{i}$ and $\mathbf{r}_{i}$ as Eq. (\ref{eq:bstep}).\;
      $\mathcal{L}_{b\_\mathrm{step}}^{MAP} += \mathcal{L}_{i, b\_\mathrm{step}}^{MAP}$;\;
  }
  Update $\Theta_{b} = \Theta_{b} - lr \cdot \nabla _ {\Theta_{b} } (-\mathcal{L}_{b\_\mathrm{step}}^{MAP}) / I $
}
\Return{$\Theta_{t}$, $\Theta_{b}$ the trained weights of the network.}\;
\caption{SGD for VBAE-soft and VBAE-hard.}
\label{alg:vae}
\end{algorithm}

\subsubsection{Unbiased Gradient Estimators}
With the stochastic latent variables reparameterized with the above strategies, unbiased gradient estimators for the objectives $\operatorname{ELM}_{\{b,t\}\_\mathrm{step}}$ w.r.t. $\phi$ can be formulated as:
\begin{equation}
\begin{aligned}
    \nabla _ {\phi} \operatorname{ELM}_{b\_\mathrm{step}} \simeq& \frac{1}{L} \sum_{l} \bigg( \nabla_{\mathbf{z}_{b}^{(l)},\ d^{(l)}} \log p(\mathbf{r}|\mathbf{v}^{(l)}) \cdot \nabla_{\phi} {[\mathbf{z}_{b}^{(l)},\ d^{(l)}]} \bigg); \\
    \nabla _ {\phi} \operatorname{ELM}_{t\_\mathrm{step}} \simeq& \frac{1}{L} \sum_{l} \bigg(\nabla_{\mathbf{z}_{t}^{(l)}} \log p(\mathbf{r}|\mathbf{v}^{(l)})\cdot \nabla_{\phi} \mathbf{z}_{t}^{(l)} \bigg) + \\
    & \frac{1}{L} \sum_{l} \bigg(\nabla_{\mathbf{z}_{t}^{(l)}} \log p(\mathbf{x}|\mathbf{z}_{t}^{(l)})\cdot \nabla_{\phi} \mathbf{z}_{t}^{(l)} \bigg),
\end{aligned}
\end{equation}
\noindent where $\simeq$ denotes that the RHS. is an unbiased estimator of the LHS. Since the variance of the gradient estimator of reparameterization trick is low, previous work has demonstrated that as long as the batch size is large enough, it suffices to take a single sample for each user for the training to converge \cite{gal2016uncertainty}. The optimization procedure of VBAE is summarized in Algorithm \ref{alg:vae} for reference.

\subsection{Prediction for New Users} \label{sec:prediction}

After defining the generative and inference networks of VBAE, our discussion shifts towards how to predict new relevant items for users given their observed ratings $\mathbf{r}_{o}$ and noisy features $\mathbf{x}$ with trained VBAE. For a user, we first calculate the mean of the collaborative embedding $\boldsymbol{\mu}_{b}$ and the bandwidth $\alpha$ from the ratings via the collaborative inference network, and the mean of the feature embedding $\boldsymbol{\mu}_{t}$ from user features via the feature inference network. The user latent variable $\mathbf{v}$ can then be approximated as:
\begin{equation}
    \mathbb{E}\left[V | \mathbf{r}_{o}, \mathbf{x} \right] \approx \mathbb{E}\left[Z_{b} | \mathbf{r}_{o}\right] + \bar{d} \cdot \mathbb{E}\left[ Z_{t}  | \mathbf{x}\right] = \boldsymbol{\mu}_{b} + \bar{d} \cdot  \boldsymbol{\mu}_{t}.
\end{equation}
\noindent To avoid random behavior of the model in the testing phase, for VBAE-hard, we set $\bar{d}$ to a fixed sample from $Bernoulli(\alpha)$ to determine whether or not to incorporate user feature information in $\boldsymbol{\mu}_{t}$ to support the recommendation. For VBAE-soft, we use the mean of the Beta channel, \textit{i.e.}, $\alpha$, as $\bar{d}$. Finally, we calculate the multinomial probabilities of the remaining items from $\mathbf{v}$ via $\operatorname{gen}_{b}$ as:
\begin{equation}
\label{eq:rank}
    \mathbb{E}\left[R | \mathbf{r}_{o}, \mathbf{x}\right]=\mathbb{E}\left[\operatorname{Gen}_{b}(V) | \mathbf{r}_{o}, \mathbf{x}\right] \approx \operatorname{Gen}_{b}(\mathbb{E}\left[V | \mathbf{r}_{o}, \mathbf{x}\right]),
\end{equation}
\noindent where $\approx$ is due to the non-linearity of $\operatorname{Gen}_{b}$, and the estimated logits of unobserved items are sorted to get the final ranked list of items for recommendations.

\section{Empirical Study}
\label{sec:emp}

\begin{figure*}
\centering
\includegraphics[width=0.92\linewidth]{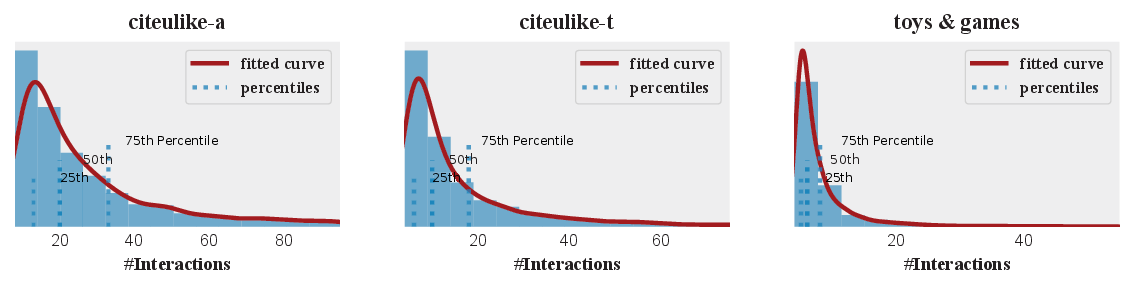}
\caption{The distribution of interaction density for different users in \textit{citeulike-a}, \textit{citeulike-t}, and Amazon \textit{toys \& games} datasets. The red curves illustrate the estimated probability density functions and the light blue dashed vertical lines shows the percentiles. (The maximum interaction value is cut off to 100, 80, 60 for the datasets for better illustration effect)}
\label{fig:density}
\end{figure*}

\begin{table}[]
\centering
\caption{Attributes of \textit{citeulike-a}, \textit{citeulike-t} and \textit{toys \& games} datasets after preprocessing. In the table, \% interactions refers to the density of the
rating matrix, max/min/avg/std \#visits refer to the statistics
of the number of item for which users provide implicit feedback in the datasets.}
\label{tab:dataset}
\begin{tabular}{lccc}
\toprule
               & {\textbf{citeulike-a}} & {\textbf{citeulike-t}} & {\textbf{toys \& games}}\\
\midrule 
\# users        & 5,551            & 5,031       & 14,706\\
\# items        & 16,980           & 21,133      & 11,722\\
\% interactions & 0.217\%          & 0.114\%     & 0.072\%\\
max/min \#visits & 403/10          & 1,932/5     & 546/5\\
avg$\pm$std \#visits & $12 \pm 12$  & $24 \pm 58$ & $8 \pm 8$ \\
\midrule
\# features & 8,000 & 20,000 & 8,000 \\
\bottomrule
\end{tabular}
\end{table}

In this section, we analyze the extensive experiments we conducted on three real-world datasets to demonstrate the effectiveness of VBAE for hybrid recommender systems.

\subsection{Datasets} \label{sec:data}

We use three real-world datasets to evaluate the models. Two of the datasets, \textit{citeulike-a} \cite{wang2011collaborative} and \textit{citeulike-t} \cite{wang2013collaborative} are from CiteULike, where scholars can add academic articles they are interested in to their libraries such that new relevant articles can be automatically recommended. The third dataset, \textit{toys \& games}, is collected by \cite{he2016ups} from Amazon\footnote{\url{https://nijianmo.github.io/amazon/index.html}}. In preprocessing, we randomly split users by the ratio of 8:1:1 for training, validation, and testing. For each user, 80\% of the interactions are selected as the observed interactions to learn the user collaborative embedding and the channel's bandwidth, and the remaining 20\% are held out for testing. The user profiles are built from the features of their interacted items. We represent each article in the \textit{citeulike} datasets by the concatenation of its title and abstract, and each item in \textit{toys \& games} by combining the reviews it received at Amazon. We then select discriminative words according to the tf-idf values and normalize the word counts over their maximum occurrences in all items. Finally, we calculate the element-wise maximum of the normalized word counts of interacted items as user features. Table \ref{tab:dataset} summarizes the detail of the datasets after preprocessing. Moreover, we also illustrate the distribution of interaction density for different users in Fig. \ref{fig:density}. From Fig. \ref{fig:density} we can find that the interaction density distribution clearly demonstrates a long-tail characteristic, which reflects the unevenness of sufficiency level of collaborative information for users in all three datasets.

\subsection{Evaluation Metrics}

Two ranking-based metrics are used to evaluate the recommendation performance: Recall@$M$ and truncated normalized discounted cumulative gain (NDCG@$M$). We do not use the precision metric because rating matrices in all three datasets record implicit feedbacks, where a zero entry does not necessarily imply that the user lacks interests, but it may also indicate that the user is not aware of the item's existence \cite{hu2008collaborative}. For a user $i$, we first obtain the rank of the hold-out items by sorting their multinomial probabilities. If we denote the item at rank $r$ by $j(r)$ and the set of hold-out items for user $i$ by $\mathcal{J}_{i}$, Recall@$M$ is calculated as:
\begin{equation}
\operatorname{Recall} @ M(i)=\frac{\sum_{r=1}^{M} \mathbb{I}\left[j(r) \in \mathcal{J}_{i}\right]}{\min \left(M,\left|\mathcal{J}_{i}\right|\right)},
\end{equation}
\noindent where $\mathbb{I}$ in the numerator is the indicator function, and the denominator is the minimum of $M$ and the number of hold-out items. Recall@$M$ has a maximum of 1, which is achieved when all relevant items are ranked among the top $M$ positions. Truncated discounted cumulative gain (DCG@$M$) is computed as
\begin{equation}
    \operatorname{DCG} @ M(i)=\sum_{r=1}^{M} \frac{2^{\mathbb{I}\left[j(r) \in \mathcal{J}_{i}\right]}-1}{\log (r+1)},
\end{equation}
\noindent which, instead of uniformly weighting all positions, introduces a logarithm discount function over the ranks where larger weights are applied to recommended items that appear at higher ranks \cite{wang2013theoretical}. NDCG@$M$ is calculated by normalizing the DCG@$M$ to [0, 1] by the ideal DCG@$M$ where all relevant items are ranked at the top.

\subsection{Implementation Detail} \label{sec:imple}

Since the datasets we consider vary both in their scale and scope, we select the structure and the hyperparameters of VBAE based on evaluation metrics on validation users through grid search\footnote{Due to space limit, please refer to the codes for the searched optimal hyperparameters and model architecture for the three datasets.}. In VBAE, sigmoid is used as both intermediate and output activations. The weights of the inference network are tied to the generation network the same way as \cite{li2017collaborative} to more effectively learn representations of user features. Specifically, to avoid the component collapsing problem where the inferred bandwidth for all users is identical, batch normalization \cite{ioffe2015batch} is applied to the L2-norm of the latent feature representations such that they have zero mean and unit variance; in addition, a larger decay rate is applied to the weights of the dense layer for bandwidth inference for regularization. We first layerwise pretrain the user feature network as the initial point for VBAE. Then we iteratively train the collaborative network ($b$\_step) and the user feature network ($t$\_step) for 100 epochs. Adam is used as the optimizer with a mini-batch of 500 users. We randomly split the datasets into ten train/val/test splits as described in Section \ref{sec:data}. For each split, we keep the model with the best NDCG@100 on the validation users and report the test metrics averaged on ten splits of the datasets.

\subsection{Baselines} \label{sec:baselines}

In this section, we compare the proposed VBAE with the following state-of-the-art collaborative and hybrid recommendation baselines to demonstrate its effectiveness:

\begin{itemize}
    \item \textbf{FM} (Factorization Machine) is a widely employed algorithm for hybrid recommendation with sparse inputs \cite{guo2017deepfm}. We use Bayesian parameter search as suggested in \cite{dacrema2019we} to find the optimal hyperparameters and] loss function on the validation users.
    
    \item \textbf{CTR} \cite{wang2011collaborative} learns the topics of item content via latent Dirichlet allocation (LDA) and couples it with probabilistic matrix factorization  (PMF) for collaborative filtering. We find the optimal hyperparameters $a$, $b$, $\lambda_{u}$, $\lambda_{v}$ and latent dimension $K$ through grid search.
    
    \item \textbf{CDL} \cite{wang2015collaborative} replaces the LDA in CTR with a stacked Bayesian denoising auto-encoder (SDAE) \cite{vincent2010stacked} to learn the item content embeddings in an end-to-end manner. We set the mask rate of SDAE to 0.3 and search its architecture the same as VBAE.
    
    \item \textbf{CVAE} \cite{li2017collaborative} further improves over the CDL by utilizing a VAE in place of the Bayesian SDAE, where a self-adaptive Gaussian noise is introduced to corrupt the latent item embeddings instead of corrupting the input features with zero masks.
    
    \item \textbf{Multi-VAE} \cite{liang2018variational} breaks the linear collaborative modeling bottleneck of PMF by using a VAE with multinomial likelihood to capture the user collaborative information in ratings for recommendations.
    
    \item \textbf{CoVAE} \cite{chen2018collective} utilizes the non-linear Multi-VAE as the collaborative backbone and incorporates item feature information by treating their co-occurrences as pseudo training samples to collectively train the Multi-VAE with item features. 
    
    \item \textbf{CondVAE} \cite{pang2019novel} builds a user conditional VAE based on Multi-VAE where user features are used as the conditions. We extend the original CondVAE by replacing the categorical user features with the ones we build from the interacted items, which we find could have a better performance on all the datasets.
    
    \item \textbf{DICER} \cite{zhang2020content} is an item-oriented auto-encoder (IAE)-based recommender system that learns content-disentangled item representations from their user ratings to achieve more robust recommendations.
    
    \item \textbf{RecVAE} \cite{shenbin2020recvae} improves over the Multi-VAE by designing a new encoder architecture with a composite prior for user collaborative latent variables, which leads to a more stable training procedure.
    
\end{itemize}

\begin{table}[]
\caption{Comparisons between VBAE and various baselines. We report the metrics averaged on ten splits of the datasets. The best method is highlighted in \textbf{bold}, where the best method in each part is marked with underlines.}

\begin{subtable}[t]{\columnwidth}
\centering
\caption{citeulike-a}\label{tab:a_base}
\begin{tabular}{lccc}
\makebox[0.17\textwidth][c]{{}}& \makebox[0.17\textwidth][c]{{Recall@20}} & \makebox[0.17\textwidth][c]{{Recall@40}} & \makebox[0.17\textwidth][c]{{NDCG@100}} \\
\toprule                                     
VBAE-soft & \textbf{0.299}  & \textbf{0.376}  & \textbf{0.296}  \\
VBAE-hard & 0.293  & 0.373  & 0.294  \\
\midrule
FM   & 0.231  & 0.312  & 0.238 \\
CTR  & 0.169  & 0.250  & 0.190  \\
CDL  & 0.209  & 0.295  & 0.226  \\
CVAE & \underline{0.236}  & \underline{0.334}  & \underline{0.247}  \\
\midrule
Multi-VAE  & 0.261 & 0.346 & 0.265 \\
RecVAE & 0.265 &  0.354 &  0.269   \\
CoVAE & 0.247 & 0.338 & 0.260 \\
CondVAE & 0.274 & 0.359 & \underline{0.275} \\
DICER &  \underline{0.279} & \underline{0.363} & 0.272\\
\bottomrule
\end{tabular}
\end{subtable}

\bigskip

\begin{subtable}[t]{\columnwidth}
\centering
\caption{citeulike-t}\label{tab:t_base}
\begin{tabular}{lccc}
\makebox[0.17\textwidth][c]{{}}& \makebox[0.17\textwidth][c]{{Recall@20}} & \makebox[0.17\textwidth][c]{{Recall@40}} & \makebox[0.17\textwidth][c]{{NDCG@100}} \\
\toprule
VBAE-soft & \textbf{0.227} & 0.306 & 0.190 \\
VBAE-hard & 0.223 & \textbf{0.308} &  \textbf{0.193} \\
\midrule
FM & 0.154 & 0.224 & 0.135 \\
CTR   & 0.132 & 0.189 & 0.118 \\
CDL   & 0.200 & 0.271 & 0.168 \\
CVAE  & \underline{0.216} & \underline{0.294} & \underline{0.181} \\
\midrule
Multi-VAE & 0.168 & 0.247 & 0.139 \\
RecVAE & 0.177 & 0.251 & 0.148 \\
CoVAE & 0.194 & 0.257 & 0.167 \\
CondVAE & \underline{0.215} & 0.279 & \underline{0.172} \\
DICER & 0.203 & \underline{0.283} & 0.161\\
\bottomrule
\end{tabular}
\end{subtable}

\bigskip

\begin{subtable}[t]{\columnwidth}
\centering
\caption{toys \& games}\label{tab:v_base}
\begin{tabular}{lccc}
\makebox[0.17\textwidth][c]{{}}& \makebox[0.17\textwidth][c]{{Recall@20}} & \makebox[0.17\textwidth][c]{{Recall@40}} & \makebox[0.17\textwidth][c]{{NDCG@100}} \\
\toprule             
VBAE-soft & \textbf{0.145} & \textbf{0.196} & \textbf{0.107} \\
VBAE-hard & 0.144 & 0.193 & 0.105 \\
\midrule
FM   & 0.088 & 0.121 & 0.062 \\
CTR  & 0.124  & 0.179  & 0.089  \\
CDL  & 0.133  & 0.181  & 0.092  \\
CVAE & \underline{0.139} & \underline{0.188} & \underline{0.094} \\
\midrule
Multi-VAE & 0.114 & 0.157 & 0.082 \\
RecVAE & 0.110 & 0.154 & 0.077 \\
CoVAE & 0.120 & 0.174 & 0.085 \\
CondVAE & \underline{0.132} & \underline{0.180} & \underline{0.094} \\
DICER & 0.127 & 0.172 & 0.092 \\
\bottomrule
\end{tabular}
\end{subtable}

\label{tab:results}
\end{table}

Comparison results between the two VBAE models and the baselines are summarized in Table \ref{tab:results}. 

\subsection{Comparison Analysis}

As it can be seen, Table \ref{tab:results} comprises of three parts. The middle part shows four hybrid baselines with linear collaborative filtering module, \textit{i.e.,} matrix factorization (MF). Generally, the performance improves with the increase of representation ability of the utilized item content embedding model. Specifically, CVAE, which uses VAE to encode item content information into Gaussian variables, performs consistently better than CDL and CTR on all three datasets. However, we also observe that simple methods such as FM can outperform some of the deep learning-based baselines (\textit{e.g.}, CDL on \textit{citeulike-a} datasets) when their parameters are systematically searched with a Bayesian optimizer, which is consistent with the findings in \cite{dacrema2019we}. 

The bottom part shows baselines that utilize deep neural networks (DNNs) as the collaborative module. Multi-VAE, RecVAE can capture non-linear similarities among users, and they improve almost consistently over the linear hybrid baselines when the datasets are comparatively dense (\textit{e.g.,} the \textit{citeulike-a} dataset), even if they do not use user or item side information. When the datasets get sparser, however, they cannot perform on par with the PMF-based hybrid recommenders augmented with item side information due to lack of sufficient collaborative information. Moreover, we find that CoVAE does not consistently outperform Multi-VAE, which could suggest that the item feature occurrences do not necessarily imply user co-purchases. Utilizing user feature embeddings as the condition for the user collaborative embeddings, CondVAE achieves the best performance among all UAE-based baselines on the denser \textit{citeulike} datasets and performs on par with CVAE on the sparser Amazon \textit{toys \& games} dataset. As an IAE-based recommender, DICER shows clear merits when the dataset has a large item-to-user ratio (\textit{e.g.,} the \textit{citeulike-a} and \textit{citeulike-t} datasets). The reason could be that for IAE-based recommenders, the number of training samples is proportional to the number of items, whereas the number of trainable weights is proportional to the number of users, and therefore a large item-to-user ratio ensures sufficient training samples to guarantee the model generalization ability.

Simultaneously addressing the uncertainty in user ratings and noise in user features, VBAE-soft and VBAE-hard out-performs all baselines on all three datasets. Although the Bayesian SDAE in CDL, VAE in CVAE, RecVAE also have the denoising ability in that they corrupt the item features, latent item embeddings, or latent user embeddings via masked noise or self-adaptive Gaussian noise, the noise they address is not recommendation-oriented and is therefore inevitably low-level. However, high-level and personalized noise (information that is not relevant to the recommendation purpose) pervasively exists in recommendation tasks, which cannot be addressed by these models. In contrast, through the introduction of a user-dependent channel variable, VBAE actively decides how much information should be accessed from user features based on information already contained in user ratings. This ensures the quality of personalized recommendations when ratings are sparse while improving the model generalization ability by avoiding unnecessary dependence on noisy user features when the collaborative information is sufficient.

\subsection{Ablation Study on the User-dependent Channel} \label{sec:channels}

In this section, we further demonstrate the effectiveness of the established information regulation mechanism in VBAE by answering the following two research questions: 

\noindent \quad \textbf{RQ1:} How do VBAE-hard and VBAE-soft perform compared to VBAE-like models, which, instead of explicitly considering the personal difference in the sufficiency level of collaborative information and the pervasive noise in user features, treat the fusion of user feature information into user latent variables as a fixed procedure for all the users. 

\noindent \quad \textbf{RQ2:} How does the dynamic channel perform compared with the vanilla attention mechanism, and how well does the inferred bandwidth correspond to the deficiency of collaborative information? The answer to this question shows the effectiveness of the proposed quantum-inspired collaborative uncertainty measurement to distinguish users with varied sufficiency levels of collaborative information.

\subsubsection{Comparisons with Ablation Baselines}

To answer \textbf{RQ1}, we design the following five baseline models as ablation studies: 

\begin{itemize}
    \item \textbf{DBAE-pass}  uses an "all-pass" channel to connect the user collaborative and feature networks, where all the information in user feature embeddings are losslessly transferred to the corresponding user latent variables irrespective of the individual difference in the sufficiency level of collaborative information; 
    
    \item \textbf{DBAE-stop} uses a "stop" channel that indiscriminately blocks the auxiliary feature information  for all users, and only the collaborative information is exploited to calculate user latent variables. 
    
    \item \textbf{VAE-concat} concatenates user ratings and features as the inputs to Multi-VAE and reconstructs the ratings for recommendations. The fusion can be viewed as learning a fixed weighted combination of user ratings and features for all users.

    \item \textbf{VAE-attn} uses the vanilla attention mechanism to fuse the user collaborative and feature embeddings. The attention weights for the rating and feature embeddings are calculated by $\alpha_{\{b, t\}} = e^{\mathbf{w}_{\{b,t\}} \cdot \mathbf{h}_{\{b,t\}}} / \sum_{\{b,t\}} e^{\mathbf{w}_{\{b,t\}} \cdot \mathbf{h}_{\{b,t\}}}$, where $\mathbf{w}_{\{b,t\}}$ are trainable embedding vectors.

    \item \textbf{VBAE-nn} uses a simple neural network to calculate the bandwidth $\alpha$ from the user collaborative embeddings $\mathbf{h}_{b}$. The structure of the network is determined by grid search. This model is used to further verify the effectiveness of the proposed quantum-inspired semantic-uncertainty inference strategy.
\end{itemize}

\begin{table}[]
\caption{Comparisons between VBAE and various baselines. The best method is highlighted in \textbf{bold}, where the best method in each part is marked with underlines.}
\begin{subtable}[t]{\columnwidth}
\centering
\caption{citeulike-a}\label{tab:a_channel}
\begin{tabular}{lcccc}
          & Recall@20 & NDCG@100 & Bandwidth & PCC\\
\toprule
VBAE-soft & \textbf{0.299} & \textbf{0.296} & 0.543 {$\pm$ 0.054} & -0.898\\
VBAE-hard & 0.293 & 0.294 &  0.812 {$\pm$ 0.048} & -0.901 \\
\midrule
VBAE-nn (soft) & \underline{0.292} & \underline{0.291}  & 0.610 $\pm$ 0.046 & -0.871 \\
VBAE-nn (hard) & 0.289 & 0.287  & 0.925 $\pm$ 0.031 & -0.883 \\
\midrule
DBAE-stop & 0.263 & 0.269  & 0.000 {$\pm$ 0.000} & N/A \\
DBAE-pass & \underline{0.287} & 0.285  & 1.000 {$\pm$ 0.000} & N/A \\
VAE-attn   & 0.282 & \underline{0.286} &  N/A & N/A \\
VAE-concat & 0.274 & 0.280 &  N/A & N/A \\
\bottomrule
\end{tabular}
\end{subtable}

\bigskip

\begin{subtable}[t]{\columnwidth}
\centering
\caption{citeulike-t}\label{tab:t_channel}
\begin{tabular}{lcccc}
          & Recall@20 & NDCG@100 & Bandwidth & PCC\\
\toprule
VBAE-soft & \textbf{0.227} & 0.190 & 0.546 {$\pm$ 0.050 } & -0.887\\
VBAE-hard & 0.223 & \textbf{0.193} & 0.805 {$\pm$ 0.035 } & -0.910 \\
\midrule
VBAE-nn (soft) & \underline{0.221} & \underline{0.186}  & 0.615 {$\pm$ 0.043} & -0.882 \\
VBAE-nn (hard) & 0.218 & 0.184  & 0.886 {$\pm$ 0.037} & -0.903 \\
\midrule
DBAE-stop & 0.170 & 0.142 & 0.000 {$\pm$ 0.000} & N/A \\
DBAE-pass & 0.212 & \underline{0.178} & 1.000 {$\pm$ 0.000} & N/A \\
VAE-attn   & \underline{0.219} & 0.177 &  N/A & N/A \\
VAE-concat & 0.215 & 0.172 &  N/A & N/A \\
\bottomrule
\end{tabular}
\end{subtable}

\bigskip

\begin{subtable}[t]{\columnwidth}
\centering
\caption{toys \& games}\label{tab:v_channel}
\begin{tabular}{lcccc}
          & Recall@20 & NDCG@100 & Bandwidth & PCC\\
\toprule
VBAE-soft & \textbf{0.145} & \textbf{0.107} & 0.560 {$\pm$ 0.057} & -0.803\\
VBAE-hard & 0.144 & 0.105 & 0.829 {$\pm$ 0.031} & -0.825 \\
\midrule
VBAE-nn (soft) & \underline{0.139} & \underline{0.103}  & 0.643 {$\pm$ 0.052} & -0.814 \\
VBAE-nn (hard) & 0.137 & 0.099  & 0.901 {$\pm$ 0.036} & -0.828 \\
\midrule
DBAE-stop & 0.119 & 0.088 & 0.000 {$\pm$ 0.000} & N/A\\
DBAE-pass & 0.135 & 0.094 & 1.000 {$\pm$ 0.000} & N/A\\
VAE-attn   & \underline{0.136} & \underline{0.097} &  N/A & N/A \\
VAE-concat & 0.132 & 0.095 &  N/A & N/A \\
\bottomrule
\end{tabular}
\end{subtable}

\flushleft
\textbf{\small {The $\pm$ in the bandwidth column denotes the std of users' bandwidth averaged over ten splits of the datasets.}}

\label{tab:results_channel}
\end{table}

\begin{figure*}
\centering
\includegraphics[width=0.95\linewidth]{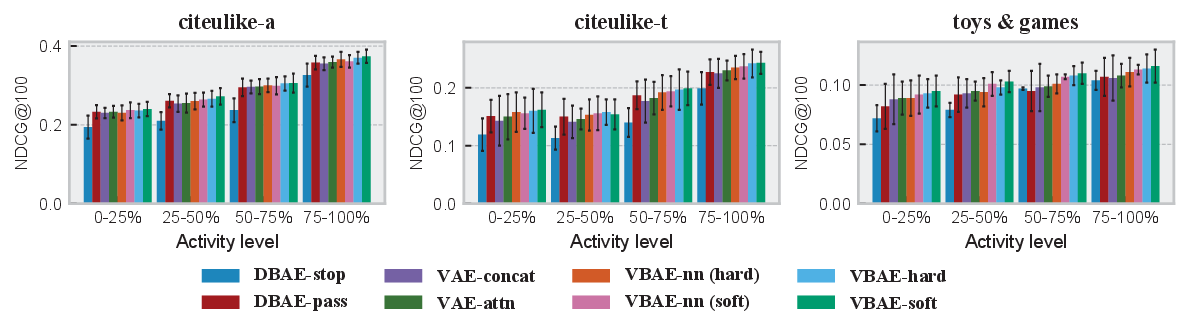}
\caption{NDCG@100 breakdown for users with different interaction activity levels measured by the number of historically interacted items. The error bar represents the standard deviation across ten random dataset splits.}
\label{fig:breakdown}
\end{figure*}

\noindent The comparison results are listed in Table \ref{tab:results_channel}. From Table \ref{tab:results_channel} we can find that, among the models that we draw comparisons with, DBAE-stop performs the worst on all the datasets. Since DBAE-stop can be viewed as an altered version of Multi-VAE \cite{li2017collaborative} where the L2-normalization is applied on the hidden representations rather than the input ratings, this confirms the previous finding that hybrid recommendation methods augmented with feature information usually perform better than collaborative-filtering-based methods when interactions are sparse \cite{wang2015collaborative, li2017collaborative}. Comparatively, DBAE-pass is more difficult to beat, since the deficiency of collaborative information for users with sparse interactions makes the auxiliary user feature information valuable even if they are noisy. Still, two VBAE-based methods perform better on all three datasets, which demonstrates that dynamically regulating user feature information allowed to be accessed by user latent variables can indeed improve model generalization. VAE-concat uses a dense layer to learn a weighted combination of user ratings and features, which may be over-parameterized and prone to overfitting when datasets are sparse. Moreover, the weights are shared among all users, which fails to distinguish users with different sufficiency levels of collaborative information. The superiority of VBAE-hard and VBAE-soft to VBAE-pass and VAE-concat indicates that for users with more informative interactions (\textit{i.e.,} denser and overlapped more heavily with the interactions of other users), the collaborative information is \textit{per se} very reliable for recommendations, and the noise introduced by the fusion of user features may outweigh the useful information and degenerate the recommendation performance. Finally, the favorable comparisons of VBAE-soft and VBAE-hard to VAE-attn and two VBAE-nn models demonstrate the effectiveness of the proposed quantum-inspired semantic-uncertainty measurement of collaborative information, where the uncertainty inferred via the length of hidden collaborative embeddings can better reflect the relative importance between user ratings and features.

The explanation for the general superiority of VBAE-soft over VBAE-hard could be that VBAE-soft uses a Beta channel variable with its variance fixed to a small value, where user feature embeddings are stably and smoothly discounted based on the bandwidth inferred from user ratings. In contrast, the Bernoulli channel in VBAE-hard determines whether or not to access user features with the bandwidth as the access probability, which may be coarse in granularity and makes the training dynamic less stable than the Beta channel in VBAE-soft. However, we also observe an exception that VBAE-hard has better NDCG@100 than VBAE-soft on \textit{citeulike-t} dataset. Based on the attributes of the three datasets in Table \ref{tab:dataset}, we speculate the reason could be that since the user features in \textit{citeulike-t} are established by calculating the normalized word counts for words with top 20,000 tf-idf values instead of 8,000 (for both \textit{citeulike-a} and \textit{toys \& games} datasets), the features in \textit{citeulike-t} dataset are nosier, which makes completely discarding the noisy feature information in VBAE-hard more advantageous compared with discounting it through the soft channel.

\subsubsection{NDCG Breakdown w.r.t. User Activity Levels}

To further investigate the effectiveness of the user-dependent channel to fuse the collaborative and feature information for users with different activity levels, we divide test users into quartiles and report the averaged NDCG@100 on each user group in Fig. \ref{fig:breakdown}. When comparing with DBAE-stop, we mainly focus on users with low activity levels, since for these users, DBAE-stop accesses no user feature information while VBAE-hard and VBAE-soft infer a large channel bandwidth that allows more information to be accessed from user features. The leftmost bar group in Fig. \ref{fig:breakdown} shows that VBAE-hard and VBAE-soft significantly outperform DBAE-stop on all three datasets. The result confirms that incorporating auxiliary user feature information can alleviate the uncertainty of user collaborative embeddings and improve recommendation performance when ratings are extremely sparse, even if user features are generally noisy. When comparing with DBAE-pass, on the other hand, we focus on users with high activity levels. Although for these users, VBAE-hard and VBAE-soft access less information from user features than DBAE-pass, the rightmost bar group of Fig. \ref{fig:breakdown} shows that NDCG@100 improves consistently for these users. This further demonstrates that for users with dense interactions, the collaborative information in the ratings is \textit{per se} very reliable for recommendations, and the noise introduced by the fusion of user features may outweigh the useful information and lower the recommendation performance. The improvement is more significant on \textit{citeulike-t} and \textit{toys \& games} datasets. Recall that Table 1 shows that users in these two datasets span a wider spectrum in their activity levels. This indicates that the reliability of user collaborative embeddings varies drastically for users in both datasets. In such a case, the channel can better distinguish these users and allocate for each user a suitable budget for user feature information when calculating the user latent variables for recommendations. 

\subsubsection{Statistical Analysis of the Bandwidth}

To answer \textbf{RQ2}, we calculate several statistics of the inferred bandwidth for all test users: its averaged value, its user variability, and its Pearson correlation coefficient (PCC) with interaction density, and report the results in Table \ref{tab:results_channel}. Table \ref{tab:results_channel} shows that the bandwidth inferred through the proposed quantum-inspired collaborative information uncertainty measurement tends to vary across users with different levels of rating density. Moreover, the bandwidth has an over -0.8 PCC with the interaction density on all three datasets. Such results indicate that the channel in VBAE-hard and VBAE-soft can well distinguish users with different sufficiency levels of collaborative information and can dynamically control the extra amount of information that needs to be accessed from user features based on inferred bandwidth, which more convincingly demonstrates the effectiveness of the user-dependent channel in VBAE-hard and VBAE-soft. In addition, the average bandwidth of VBAE-hard is significantly larger than that of VBAE-soft on all three datasets. The reason could be that a larger bandwidth for VBAE-hard is conducive to maintaining the stability of the Bernoulli channel in the training phase.

\subsection{Discussions of Broader Impact of VBAE}

Although we demonstrate the effectiveness of VBAE by its application in recommender systems, VBAE is a general framework that is applicable to any heterogeneous information system where one information source is comparatively reliable but could be missing, whereas another information source is abundant but is susceptible to noise. One typical example is the "audio-assisted action recognition in the dark" task \cite{xu2021arid}, which aims to detect actions in under-illuminated videos. In the task, the visual information is more reliable for action prediction but could be missing due to bad illumination, whereas the audio track always accompanies the video but may contain lots of irrelevant information (\textit{e.g.,} background music) for the action recognition purpose. To apply VBAE to these new tasks, the only mandatory change required is to design a suitable per data point uncertainty measurement of the primary information source to dynamically decide the information allowed to be accessed from the second source, so that the model will not overfit on noise in the auxiliary modality.

Moreover, VBAE can be readily generalized to more complex heterogeneous information systems where more than two modalities exist. One simple strategy for the adaptation is to identify one modality with the highest reliability and the most severe data missing problem as the primary modality (such a modality probably exists because high reliability of an information source usually means it is difficult or expensive to obtain) and the remaining modalities as the auxiliary modalities. We first fuse the embeddings of auxiliary modalities with commonly used multimodal fusion techniques such as concatenation or product-of-experts principle. Afterward, a dynamic channel similar to the one used in VBAE can be established to dynamically regulate the fusion of information in the auxiliary modalities based on the information sufficiency level of the primary modality. Therefore, we speculate that VBAE could have a broader impact on areas of data mining other than hybrid recommender systems discussed in this article.

\section{Conclusions} 
\label{sec:con}
In this article, we develop an information-driven generative model called collaborative variational bandwidth auto-encoder (VBAE) to address uncertainty and noise problems associated with two heterogeneous sources, \textit{i.e.,} ratings and user features, in hybrid recommender systems. In VBAE, we establish an information regulation mechanism to fuse the collaborative and feature information, where a user-dependent channel is introduced to dynamically control the amount of information allowed to be accessed from user features given the information already contained in user collaborative embeddings. The channel alleviates the uncertainty problem when observed ratings are sparse while improving the model generalization ability with respect to new users by avoiding unnecessary dependence on noisy user features. The effectiveness of VBAE is demonstrated by extensive experiments on three real-world datasets.

\definecolor{link}{HTML}{2D2F92}


\begin{thebibliography}{10}
\providecommand{\url}[1]{#1}
\csname url@samestyle\endcsname
\providecommand{\newblock}{\relax}
\providecommand{\bibinfo}[2]{#2}
\providecommand{\BIBentrySTDinterwordspacing}{\spaceskip=0pt\relax}
\providecommand{\BIBentryALTinterwordstretchfactor}{4}
\providecommand{\BIBentryALTinterwordspacing}{\spaceskip=\fontdimen2\font plus
\BIBentryALTinterwordstretchfactor\fontdimen3\font minus
  \fontdimen4\font\relax}
\providecommand{\BIBforeignlanguage}[2]{{%
\expandafter\ifx\csname l@#1\endcsname\relax
\typeout{** WARNING: IEEEtran.bst: No hyphenation pattern has been}%
\typeout{** loaded for the language `#1'. Using the pattern for}%
\typeout{** the default language instead.}%
\else
\language=\csname l@#1\endcsname
\fi
#2}}
\providecommand{\BIBdecl}{\relax}
\BIBdecl

\bibitem{zhang2019deep}
S.~Zhang, L.~Yao, A.~Sun, and Y.~Tay, ``Deep learning based recommender system:
  A survey and new perspectives,'' \emph{ACM Computing Surveys}, vol.~52,
  no.~1, pp. 1--38, 2019.

\bibitem{wang2014relational}
H.~Wang and W.-J. Li, ``Relational collaborative topic regression for
  recommender systems,'' \emph{IEEE Trans. Knowl. Data Eng.}, vol.~27, no.~5,
  pp. 1343--1355, 2014.

\bibitem{chen2020revisiting}
L.~Chen, L.~Wu, R.~Hong, K.~Zhang, and M.~Wang, ``Revisiting graph based
  collaborative filtering: A linear residual graph convolutional network
  approach,'' in \emph{Proc. AAAI}, vol.~34, no.~01, 2020, pp. 27--34.

\bibitem{xu2013emr}
B.~Xu, J.~Bu, C.~Chen, C.~Wang, D.~Cai, and X.~He, ``{EMR}: A scalable
  graph-based ranking model for content-based image retrieval,'' \emph{IEEE
  Trans. Knowl. Data Eng.}, vol.~27, no.~1, pp. 102--114, 2013.

\bibitem{yi2021cross}
J.~Yi, Y.~Zhu, J.~Xie, and Z.~Chen, ``Cross-modal variational auto-encoder for
  content-based micro-video background music recommendation,'' \emph{IEEE
  Trans. Multimedia}, to be published, doi: \href{http://dx.doi.org/10.1109/TMM.2021.3128254}{\color{link}10.1109/TMM.2021.3128254}.

\bibitem{dong2021dual}
J.~Dong, X.~Li, C.~Xu, X.~Yang, G.~Yang, X.~Wang, and M.~Wang, ``Dual encoding
  for video retrieval by text,'' \emph{IEEE Trans. Pattern Anal. Mach.
  Intell.}, to be published, doi: \href{http://dx.doi.org/10.1109/TPAMI.2021.3059295}{\color{link}10.1109/TPAMI.2021.3059295}.

\bibitem{sun2018attentive}
P.~Sun, L.~Wu, and M.~Wang, ``Attentive recurrent social recommendation,'' in
  \emph{Proc. SIGIR}, 2018, pp. 185--194.

\bibitem{xie2020multimodal}
J.~Xie, Y.~Zhu, Z.~Zhang, J.~Peng, J.~Yi, Y.~Hu, H.~Liu, and Z.~Chen, ``A
  multimodal variational encoder-decoder framework for micro-video popularity
  prediction,'' in \emph{Proc. WWW}, 2020, pp. 2542--2548.

\bibitem{chen2020learning}
X.~Chen, D.~Liu, Z.~Xiong, and Z.-J. Zha, ``Learning and fusing multiple user
  interest representations for micro-video and movie recommendations,''
  \emph{IEEE Trans. Multimedia}, vol.~23, pp. 484--496, 2020.

\bibitem{xu2021multi}
Y.~Xu, L.~Zhu, Z.~Cheng, J.~Li, Z.~Zhang, and H.~Zhang, ``Multi-modal discrete
  collaborative filtering for efficient cold-start recommendation,'' \emph{IEEE
  Trans. Knowl. Data Eng.}, to be published, doi: \href{http://dx.doi.org/10.1109/TKDE.2021.3079581}{\color{link}10.1109/TKDE.2021.3079581}.

\bibitem{sedhain2015autorec}
S.~Sedhain, A.~K. Menon, S.~Sanner, and L.~Xie, ``\protect{AutoRec}:
  Autoencoders meet collaborative filtering,'' in \emph{Proc. WWW}, 2015, pp.
  111--112.

\bibitem{liang2018variational}
D.~Liang, R.~G. Krishnan, M.~D. Hoffman, and T.~Jebara, ``Variational
  autoencoders for collaborative filtering,'' in \emph{Proc. WWW}, 2018, pp.
  689--698.

\bibitem{li2015deep}
S.~Li, J.~Kawale, and Y.~Fu, ``Deep collaborative filtering via marginalized
  denoising auto-encoder,'' in \emph{Proc. CIKM}, 2015, pp. 811--820.

\bibitem{wang2016collaborative}
H.~Wang, X.~Shi, and D.-Y. Yeung, ``Collaborative recurrent autoencoder:
  Recommend while learning to fill in the blanks,'' in \emph{Proc. NeurIPS},
  2016, pp. 415--423.

\bibitem{zhu2019improving}
Z.~Zhu, J.~Wang, and J.~Caverlee, ``Improving top-k recommendation via joint
  collaborative autoencoders,'' in \emph{Proc. WWW}, 2019, pp. 3483--3482.

\bibitem{kingma2014auto}
D.~P. Kingma and M.~Welling, ``Auto-encoding variational {B}ayes,'' in
  \emph{Proc. ICLR}, 2014.

\bibitem{zhang2017autosvd}
S.~Zhang, L.~Yao, and X.~Xu, ``\protect{AutoSVD++}: An efficient hybrid
  collaborative filtering model via contractive auto-encoders,'' in \emph{Proc.
  SIGIR}, 2017, pp. 957--960.

\bibitem{wu2016collaborative}
Y.~Wu, C.~DuBois, A.~X. Zheng, and M.~Ester, ``Collaborative denoising
  auto-encoders for top-{N} recommender systems,'' in \emph{Proc. WSDM}, 2016,
  pp. 153--162.

\bibitem{li2017collaborative}
X.~Li and J.~She, ``Collaborative variational autoencoder for recommender
  systems,'' in \emph{Proc. KDD}, 2017, pp. 305--314.

\bibitem{wang2015collaborative}
H.~Wang, N.~Wang, and D.-Y. Yeung, ``Collaborative deep learning for
  recommender systems,'' in \emph{Proc. KDD}, 2015, pp. 1235--1244.

\bibitem{ma2019learning}
J.~Ma, C.~Zhou, P.~Cui, H.~Yang, and W.~Zhu, ``Learning disentangled
  representations for recommendation,'' in \emph{Proc. NeurIPS}, 2019, pp.
  5711--5722.

\bibitem{hou2019explainable}
Y.~Hou, N.~Yang, Y.~Wu, and S.~Y. Philip, ``Explainable recommendation with
  fusion of aspect information,'' \emph{World Wide Web}, vol.~22, no.~1, pp.
  221--240, 2019.

\bibitem{yi2021dual}
Q.~Yi, N.~Yang, and P.~Yu, ``Dual adversarial variational embedding for robust
  recommendation,'' \emph{IEEE Trans. Knowl. Data Eng.}, to be published, doi:
  \href{http://dx.doi.org/10.1109/TKDE.2021.3093773}{\color{link}10.1109/TKDE.2021.3093773}.

\bibitem{slaney2011web}
M.~Slaney, ``Web-scale multimedia analysis: Does content matter?'' \emph{IEEE
  MultiMedia}, vol.~18, no.~2, pp. 12--15, 2011.

\bibitem{baldi2012autoencoders}
P.~Baldi, ``Autoencoders, unsupervised learning, and deep architectures,'' in
  \emph{Proc. ICML Workshop}, 2012, pp. 37--49.

\bibitem{lee2017augmented}
W.~Lee, K.~Song, and I.-C. Moon, ``Augmented variational autoencoders for
  collaborative filtering with auxiliary information,'' in \emph{Proceedings
  CIKM}, 2017, pp. 1139--1148.

\bibitem{sachdeva2019sequential}
N.~Sachdeva, G.~Manco, E.~Ritacco, and V.~Pudi, ``Sequential variational
  autoencoders for collaborative filtering,'' in \emph{Proc. WSDM}, 2019, pp.
  600--608.

\bibitem{zhang2020content}
Y.~Zhang, Z.~Zhu, Y.~He, and J.~Caverlee, ``Content-collaborative
  disentanglement representation learning for enhanced recommendation,'' in
  \emph{Proc. RecSys}, 2020, pp. 43--52.

\bibitem{mnih2008prob}
A.~Mnih and R.~R. Salakhutdinov, ``Probabilistic matrix factorization,'' in
  \emph{Proc. NeurIPS}, 2007, pp. 1257--1264.

\bibitem{vincent2010stacked}
P.~Vincent, H.~Larochelle, I.~Lajoie, Y.~Bengio, and P.-A. Manzagol, ``Stacked
  denoising autoencoders: Learning useful representations in a deep network
  with a local denoising criterion,'' \emph{J. Mach. Learn. Res.}, vol.~11, no.
  Dec, pp. 3371--3408, 2010.

\bibitem{koren2008factorization}
Y.~Koren, ``Factorization meets the neighborhood: A multifaceted collaborative
  filtering model,'' in \emph{Proc. KDD}, 2008, pp. 426--434.

\bibitem{pang2019novel}
B.~Pang, M.~Yang, and C.~Wang, ``A novel top-{N} recommendation approach based
  on conditional variational auto-encoder,'' in \emph{Proc. PAKDD}, 2019, pp.
  357--368.

\bibitem{li2020adversarial}
X.~Li, C.~Wang, J.~Tan, X.~Zeng, D.~Ou, D.~Ou, and B.~Zheng, ``Adversarial
  multimodal representation learning for click-through rate prediction,'' in
  \emph{Proc. WWW}, 2020, pp. 827--836.

\bibitem{hu2008collaborative}
Y.~Hu, Y.~Koren, and C.~Volinsky, ``Collaborative filtering for implicit
  feedback datasets,'' in \emph{Proc. ICDM}, 2008, pp. 263--272.

\bibitem{dong2017hybrid}
X.~Dong, L.~Yu, Z.~Wu, Y.~Sun, L.~Yuan, and F.~Zhang, ``A hybrid collaborative
  filtering model with deep structure for recommender systems,'' in \emph{Proc.
  AAAI}, 2017, pp. 1309--1315.

\bibitem{Goyal2020The}
A.~Goyal, Y.~Bengio, M.~Botvinick, and S.~Levine, ``The variational bandwidth
  bottleneck: Stochastic evaluation on an information budget,'' in \emph{Proc.
  ICLR}, 2020.

\bibitem{deng2018latent}
Y.~Deng, Y.~Kim, J.~Chiu, D.~Guo, and A.~Rush, ``Latent alignment and
  variational attention,'' in \emph{Proc. NeurIPS}, 2018, pp. 9712--9724.

\bibitem{blei2017variational}
D.~M. Blei, A.~Kucukelbir, and J.~D. McAuliffe, ``Variational inference: A
  review for statisticians,'' \emph{Journal of the American Statistical
  Association}, vol. 112, no. 518, pp. 859--877, 2017.

\bibitem{li2019cnm}
Q.~Li, B.~Wang, and M.~Melucci, ``\protect{CNM}: An interpretable
  complex-valued network for matching,'' in \emph{Proc. NAACL}, 2019, pp.
  4139--4148.

\bibitem{Johnson2003}
R.~A. Johnson, D.~W. Wichern \emph{et~al.}, ``Multivariate linear regression
  models,'' in \emph{Applied Multivariate Statistical Analysis}, 2002, ch.~7,
  pp. 360--417.

\bibitem{ioffe2015batch}
S.~Ioffe and C.~Szegedy, ``Batch normalization: Accelerating deep network
  training by reducing internal covariate shift,'' in \emph{Proc. ICML}, 2015,
  pp. 448--456.

\bibitem{BetaVAE}
I.~Higgins, L.~Matthey, A.~Pal, C.~Burgess, X.~Glorot, M.~Botvinick,
  S.~Mohamed, and A.~Lerchner, ``\protect{Beta-VAE}: Learning basic visual
  concepts with a constrained variational framework,'' in \emph{Proc. ICLR},
  2017.

\bibitem{DVIP}
A.~A. Alemi, I.~Fischer, J.~V. Dillon, and K.~Murphy, ``Deep variational
  information bottleneck,'' in \emph{Proc. ICLR}, 2017.

\bibitem{xie2021micro}
J.~Xie, Y.~Zhu, and Z.~Chen, ``Micro-video popularity prediction via multimodal
  variational information bottleneck,'' \emph{IEEE Trans. Multimedia}, to be
  published, doi: \href{http://dx.doi.org/10.1109/TMM.2021.3120537}{\color{link}10.1109/TMM.2021.3120537}.

\bibitem{BowmanVVDJB16}
S.~R. Bowman, L.~Vilnis, O.~Vinyals, A.~M. Dai, R.~J{\'{o}}zefowicz, and
  S.~Bengio, ``Generating sentences from a continuous space,'' in \emph{Proc.
  CoNLL}, 2016, pp. 10--21.

\bibitem{RezendeMW14}
D.~J. Rezende, S.~Mohamed, and D.~Wierstra, ``Stochastic backpropagation and
  approximate inference in deep generative models,'' in \emph{Proc. ICML},
  vol.~32, 2014, pp. 1278--1286.

\bibitem{jang2017categorical}
E.~Jang, S.~Gu, and B.~Poole, ``Categorical reparameterization with
  gumbel-softmax,'' in \emph{Proc. ICLR}, 2017.

\bibitem{joo2020dirichlet}
W.~Joo, W.~Lee, S.~Park, and I.-C. Moon, ``Dirichlet variational autoencoder,''
  \emph{Pattern Recognit.}, vol. 107, p. 107514, 2020.

\bibitem{Akash2017Auto}
A.~Srivastava and C.~Sutton, ``Autoencoding variational inference for topic
  models,'' in \emph{Proc. ICLR}, 2017.

\bibitem{gal2016uncertainty}
Y.~Gal, ``Uncertainty in deep learning,'' Ph.D. dissertation, University of
  Cambridge, 2016.

\bibitem{wang2011collaborative}
C.~Wang and D.~M. Blei, ``Collaborative topic modeling for recommending
  scientific articles,'' in \emph{Proc. KDD}, 2011, pp. 448--456.

\bibitem{wang2013collaborative}
H.~Wang, B.~Chen, and W.-J. Li, ``Collaborative topic regression with social
  regularization for tag recommendation,'' in \emph{Proc. IJCAI}, 2013, pp.
  2719--2725.

\bibitem{he2016ups}
R.~He and J.~McAuley, ``Ups and downs: {M}odeling the visual evolution of
  fashion trends with one-class collaborative filtering,'' in \emph{Proc. WWW},
  2016, pp. 507--517.

\bibitem{wang2013theoretical}
Y.~Wang, L.~Wang, Y.~Li, D.~He, W.~Chen, and T.-Y. Liu, ``A theoretical
  analysis of {NDCG} ranking measures,'' in \emph{Proc. CoLT}, 2013, pp. 1--30.

\bibitem{guo2017deepfm}
H.~Guo, R.~Tang, Y.~Ye, Z.~Li, and X.~He, ``{DeepFM}: {A} factorization-machine
  based neural network for ctr prediction,'' in \emph{Proc. IJCAI}, 2017, pp.
  1725--1731.

\bibitem{dacrema2019we}
M.~F. Dacrema, P.~Cremonesi, and D.~Jannach, ``Are we really making much
  progress? A worrying analysis of recent neural recommendation approaches,''
  in \emph{Proc. RecSys}, 2019, pp. 101--109.

\bibitem{chen2018collective}
Y.~Chen and M.~de~Rijke, ``A collective variational autoencoder for top-N
  recommendation with side information,'' in \emph{Proc. WDLRS}, 2018, pp.
  3--9.

\bibitem{shenbin2020recvae}
I.~Shenbin, A.~Alekseev, E.~Tutubalina, V.~Malykh, and S.~I. Nikolenko,
  ``\protect{RecVAE}: A new variational autoencoder for top-{N} recommendations
  with implicit feedback,'' in \emph{Proc. WSDM}, 2020, pp. 528--536.

\bibitem{xu2021arid}
Y.~Xu, J.~Yang, H.~Cao, J.~Yin, and S.~See, ``Arid: A new dataset for
  recognizing action in the dark,'' in \emph{IJCAI Workshop}, vol. 1370.\hskip
  1em plus 0.5em minus 0.4em\relax Springer Nature, 2021, p.~70.

\end{thebibliography}
\end{document}